\documentclass[letterpaper]{article} 
\usepackage{aaai2026}  
\usepackage{times}  
\usepackage{helvet}  
\usepackage{courier}  
\usepackage[hyphens]{url}  
\usepackage{graphicx} 
\usepackage{natbib}  
\usepackage{caption} 
\usepackage{footnote}
\makesavenoteenv{figure}
\makesavenoteenv{figure*}
\usepackage{pgfplots}
\pgfplotsset{compat=1.18}
\usepackage{pgf-pie}
\usepackage{tikz}
\usepackage{rotating}
\usepackage{algorithm}
\usepackage{algorithmic}
\usepackage{multirow}
\usepackage{enumitem}
\usepackage{makecell}
\usepackage{tabularx}
\usepackage{subcaption}
\usepackage{enumitem}
\usepackage[most]{tcolorbox}
\usepackage{cite}
\usepackage{amsmath,amssymb,amsfonts}
\usepackage{graphicx}
\usepackage{textcomp}

\usepackage{xcolor}
\usepackage{newfloat}
\usepackage{listings}
\DeclareCaptionStyle{ruled}{labelfont=normalfont,labelsep=colon,strut=off} 
\floatstyle{ruled}
\newfloat{listing}{tb}{lst}{}
\floatname{listing}{Listing}
\title{Scalable and Efficient Large-Scale Log Analysis with LLMs: An IT Software Support Case Study}
\author{
    Pranjal Gupta\textsuperscript{\rm 1},
    Karan Bhukar\textsuperscript{\rm 2}\thanks{This work was done while Karan Bhukar was at IBM.},
    Harshit Kumar\textsuperscript{\rm 1},
    Seema Nagar\textsuperscript{\rm 1},
    Prateeti Mohapatra\textsuperscript{\rm 1},
    Debanjana Kar\textsuperscript{\rm 1}
}
\affiliations{
    \textsuperscript{\rm 1}IBM Research, India, 
    \textsuperscript{\rm 2}Amazon, India

}

\begin{document}

\maketitle

\begin{abstract}
IT environments typically have logging mechanisms to monitor system health and detect issues. However, the huge volume of generated logs makes manual inspection impractical, highlighting the importance of automated log analysis in IT Software Support. In this paper, we propose a log analytics tool that leverages Large Language Models (LLMs) for log data processing and issue diagnosis, enabling the generation of automated insights and summaries. We further present a novel approach for efficiently running LLMs on CPUs to process massive log volumes in minimal time without compromising output quality. We share the insights and lessons learned from deployment of the tool - in production since March 2024 - scaled across 70 software products, processing over 2000 tickets for issue diagnosis, achieving a time savings of 300+ man hours and an estimated \$$15,444$ per month in manpower costs compared to the traditional log analysis practices.
\end{abstract}

\begin{links}
    \link{Code}{https://github.com/Log-Analyzer/LogAn}
    \link{Demo}{https://tinyurl.com/demo-logan}~\cite{logan-demo}
\end{links}

\section{Introduction}
~\label{sec:introduction}
In an enterprise setting, software issues encountered by customers are reported to the IT support team with two key inputs: a) a log dump that provides system activity, errors, etc, b) problem description - a description of their understanding of the issue, including the observed symptoms. Since log data volumes are huge, ranging from megabytes to gigabytes, automated log analysis tools play a vital role in IT software support by enabling the processing and extraction of key insights from this data ~\cite{logsurvey}.

Traditionally, support engineers (SEs) often rely on manual log analysis using the \texttt{grep} command to search for terms like ``error" or ``failure" or use scripts with predefined rules. However, these methods have various limitations: (a) \textbf{Coverage}: Predefined rules and simple keyword searches often fail to diagnose issues because of the diversity in log formats and terminology across different systems. (b) \textbf{Maintenance}: As software updates and new features are added, log structures often change. This makes predefined rules outdated, requiring constant maintenance. This is particularly problematic in the fast paced CI/CD environments. (c) \textbf{False Positives}: Keywords like ``error" or ``failure" may appear in non-critical contexts, leading to irrelevant entries being flagged and slowing down diagnosis. (d) \textbf{Feedback}: Rule-based systems don’t evolve, lacking the ability to learn from false positives or false negatives. This results in a static process that doesn’t improve. (e) \textbf{Correlation and Causality}: Logs are often scattered across services, making manual correlation and understanding event causality difficult, especially in large-scale systems with high log volumes.

To address these limitations, researchers have applied Large Language Models (LLMs) to log analytics, showing promising results and new opportunities.~\cite{logfit, lad, LAD_24, LAD_1_2024}.
Although LLMs excel at generalizing across natural language understanding tasks~\cite{minaee2024large}, their high resource and performance costs hinder scalable production deployment~\cite{zhou2024surveyefficientinferencelarge}.
In IT Support, this challenge is further exacerbated by the massive scale and heterogeneity of log data across applications. Moreover, resource constraints often necessitate running LLMs on CPUs rather than GPUs, further amplifying the difficulty of achieving efficiency at scale. Consequently, building a unified tool that processes enormous log volumes while enabling LLMs to generalize across diverse application domains remains a critical challenge.

This paper presents an intelligent and scalable log analytics tool that utilises LLMs to overcome the limitations of traditional rule-based approaches, efficiently processing large-scale log data on CPUs in minimal time without compromising overall quality. 
The LLM predicts three types of insights from log data: (1) Golden Signals, (2) Fault Categories, and (3) Named Entities~\cite{gupta2023learning,gupta-ner}. It then generates multiple reports based on the extracted insights, providing different perspectives to assist SEs in issue diagnosis. To enable efficient LLM inference on large scale log data, we introduce Label Broadcasting - a novel technique specifically designed for LLM powered log analysis, which enables our tool to operate entirely on CPU backends.

Notably, our tool doesn't attempt to determine the root cause(s) or remediation, which requires domain knowledge of the underlying application and access to proprietary code or documentation~\cite{Saha_2022, chen2024rootkgdnovelframeworkroot}. Instead, it operates solely on logs to highlight \textit{problematic} log lines—those containing cues of underlying issues—thereby guiding SEs toward faster, focused diagnosis compared to rule-based analysis. 


Our main contributions are: (1) We present the design, implementation and deployment of a scalable log analytics tool that generates insightful reports to reduce information overload for SEs during issue diagnosis. To the best of our knowledge, it is the first system capable of efficiently processing large scale log data using LLMs entirely on CPU while producing reports within practical time limits. (2) We propose Label Broadcasting, a novel technique that enables our log analytics tool to perform LLM inference on large-scale log data, achieving substantial resource savings without compromising inference quality. Experimental results validate its scalability and efficiency in handling high-volume data. (3) We share insights and lessons learned from deploying our tool across $70$ IBM Software Support products since March 2024. Additionally, we present a case study with a finance-domain client, demonstrating how the tool aids in effective issue diagnosis, thereby reducing the cognitive load and enhancing the overall productivity.

\begin{figure*}[!ht]
    \centering
    \includegraphics[width=\textwidth]{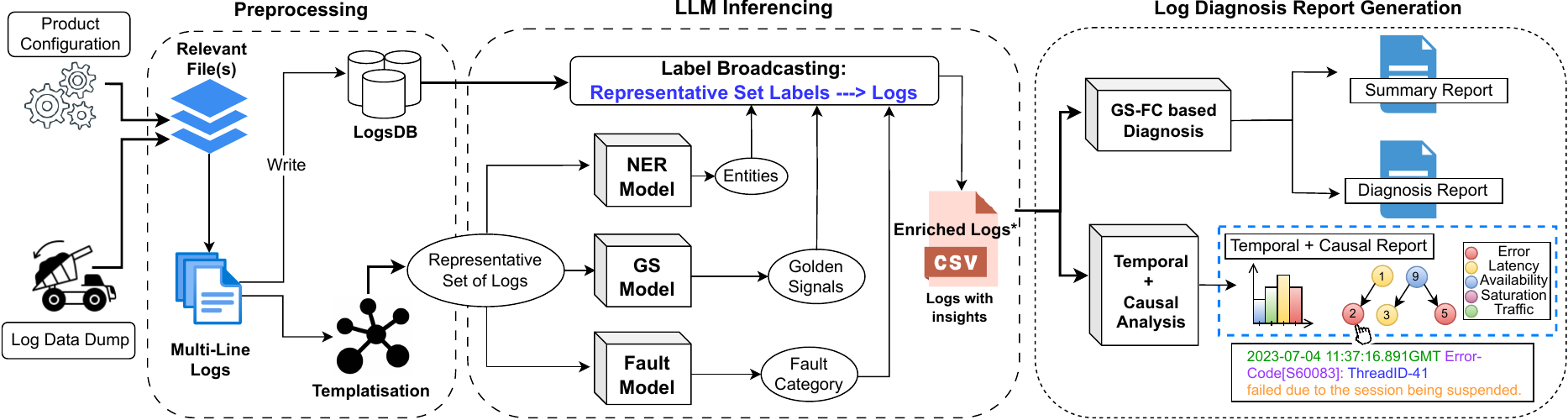}
    \caption{System Architecture of the proposed log analytics tool}
    \label{fig:logan-pipeline}
\end{figure*}
\section{System Architecture}
Analyzing massive volumes of log data is a major challenge. Applying large language models (LLMs) on every log entry is computationally expensive and slow. To overcome this, our tool comprises of three components (refer Figure~\ref{fig:logan-pipeline}) : a) \textbf{Preprocessing}: We use log templatization to group similar log lines into a smaller, representative set of clusters. This drastically reduces the data volume. b) \textbf{LLM Inferencing}: LLMs analyze the representative log lines to infer necessary labels. These labels are then broadcasted to all the log lines within their respective clusters. c) \textbf{Report Generation}: The final step is the generation of various reports, including summary, diagnosis, temporal, and causal-graph views. These reports provide a concise and actionable insights, allowing SEs to diagnose issues efficiently. The rest of the section details each of the component.
\subsection{Preprocessing}
Log data is often scattered across multiple files, which makes analysis difficult. Our tool addresses this by first consolidating all log lines from different files into a single master file. This process goes beyond simple merging, it ensures that all log lines are sorted chronologically while preserving key information such as original file name, line number, date, timestamp etc. This is crucial for correctly identifying log entries that span multiple lines. Once consolidated, this unified log data is passed to the log templatizer, which groups similar log lines together, preparing the data for further analysis.
\subsubsection{Log Templatisation}
\label{sec:log-templatisation}

Our log analytics tool leverages log templatisation~\cite{he2017drain} to extract dynamic (\textit{variables}) and constant (\textit{templates}) parts of a log line. This component groups similar log lines into one cluster (hereby known as \textit{log cluster}).
Below is an example of a log cluster with its corresponding log template (T):

\begin{itemize}
\item[] \scalebox{0.805}{\texttt{T: \textcolor{blue}{PacketResponder} \textcolor{red}{$\langle*\rangle$} \textcolor{blue}{for block} \textcolor{red}{$\langle*\rangle$} \textcolor{blue}{terminating}}} 
\item[] \scalebox{0.75}{\texttt{L1: PacketResponder \textcolor{red}{0} for block \textcolor{red}{blk\_11} terminating}}
\item[] \scalebox{0.75}{\texttt{L2: PacketResponder \textcolor{red}{1} for block \textcolor{red}{blk\_12} terminating}}
\item[] \scalebox{0.75}{\texttt{L3: PacketResponder \textcolor{red}{2} for block \textcolor{red}{blk\_13} terminating}}
\end{itemize}

The template (T) captures the syntactic structure of the log data with variables abstracted by wildcards $\langle*\rangle$. 
Log Templatisation offers many advantages such as bringing structure to data~\cite{mahindru2021log} and volume reduction~\cite{pathak2024self}.

\subsubsection{Representative Set}
\label{sec:representative-set}
The above example highlights the repetitive nature of log data, where the template of a log line remains consistent and only value of specific variables change within a log cluster.  
Notably, this characteristic may not be present in other natural languages. After templatisation, a log line is randomly selected from each log cluster to form a \textit{representative set}. This reduced set encapsulates the essence of the log dump by clustering similar lines, thereby enabling efficient LLM inference on CPUs, as explained in the following subsection.

\subsection{LLM Inferencing}
\label{sec:LLM-Inferencing}

For each log line in the representative set, the tool predicts the associated golden signal, fault category, and named entities using fine-tuned LLMs for the log domain. Three task-specific LLMs - for Golden Signal Classification (GSC), fault category prediction (FCP), and Named Entity Recognition (NER) - were finetuned offline on few manually curated, annotated multi-domain (network, SSH, Linux, etc.) log lines in a few-shot learning setup.~\cite{gupta2023learning,gupta-ner}. During runtime, the finetuned models are used directly to infer predictions for incoming logs.
Note that the training dataset used for fine-tuning the LLMs does not include logs from products deployed in production. As a result, logs from deployed products are often unseen and entirely new to both the log analytics tool and the fine-tuned LLMs. Despite this limited exposure, the LLMs are able to process and accurately classify logs from unseen domains, demonstrating strong generalizability and robustness. 
Refer to Appendix A for details on the above tasks and their comprehensive evaluation.


\subsubsection{Label Broadcasting}
\label{sec:label-broadcasting}
To scale LLM inference for the three tasks on large-scale log data, our log analytics tool performs inference only on the representative set of the log dump. The inferred labels - including golden signals, fault categories and named entities, are \textit{broadcasted} back to all the log lines within their respective log clusters. This Label Broadcasting (LB) approach reduces the number of LLM calls required, whereas conventional LLM inference optimization techniques~\cite{donisch2024inferenceoptimizationslargelanguage} still invoke the LLM for every log line. 
LB exploits the repetitive nature of log data, while such inference optimization techniques are domain-agnostic. Moreover, LB can be combined with these techniques to achieve even further inference speedups. In production, our tool uses the LB to perform LLM inference on large-scale log data entirely on CPUs.



\subsubsection{Enriched Logs}  
After broadcasting the inferred labels to all log lines in the input dump, we obtain \textit{enriched logs}, where each line is annotated with its corresponding golden signal, fault categories, and named entities. These enriched logs are further processed to generate various reports/views.

\subsection{Log Diagnosis Report Generation}
\label{subsec:log_diagnosis}
Following LLM inference, our log analytics tool generates the following diagnostic reports based on insights from enriched logs. 



\subsubsection{Summary Report} presents a table of the \textit{enriched} representative log lines (refer Figure~\ref{fig:summary-view}) — each with its predicted golden signals, fault categories, and named entities — along with the frequency of its occurrence in increasing order. The rarest one are the most important one and needs attention and therefore at the top. 
By using this approach, we've found that the tool can reduce the data volume by up to 90\%, since most log lines are informational. For easier analysis, one can filter the report by golden signals, fault categories, or named entities, which are color-coded for better readability.

\subsubsection{Diagnosis Report} presents a chronologically ordered set of relevant log windows with user-configurable granularity (e.g., 30s, 1m). A window is deemed relevant if it contains at least one erroneous or faulty signal identified through GSC or FCP via LLM inference.
Unlike the conventional methods that display the complete log dataset indiscriminately, the Diagnosis Report shows relevant log windows - a capability absent in existing solutions.
By inspecting this report, SEs can conduct pre/post analysis for issue diagnosis.
Additionally, it offers a keyword-based search interface on named entities, providing a more effective alternative to manual grep commands or predefined rules. For example, logs can be filtered based on customer-provided problem descriptions or entity types such as Error-Code. (refer Figure 11 in Appendix D) 



\subsubsection{Temporal Trend}
Since log dumps may span from days to months or even years, identifying the exact date and time of an issue often requires considerable effort, especially when such information is not explicitly provided. To mitigate this challenge, the Temporal Trend report (Figure~\ref{fig:temporal-trend-view}) visualizes the distribution of golden signals over time, thereby helping SEs track the fault progression, revealing their characteristics and severity. By analyzing spikes in erroneous signals, they can quickly understand the issue’s progression and the interplay of contributing factors.

\subsubsection{Causal Graph View}

Modern IT systems comprise numerous interdependent services, where a single underlying issue can trigger a cascade of related failures. These failures often manifest as disparate log lines scattered across different components, making manual issue diagnosis cognitively demanding for SEs. To overcome this, our log analytics tool builds a causal graph that captures the cause-effect relationships among offending/problematic representative log lines.
It captures the sequence of frequently co-occurring errors, enabling SEs to prioritize investigation and trace the origins of cascading faults. To build the causal graph, our log analytics tool utilizes enriched logs (via LLM inference) along with their associated log clusters (refer Log Templatisation section). 
A multi-variate time series is then created, representing the count occurrence of each log template within a defined time interval window.
We apply a statistical causal inference method based on Granger causality~\cite{granger1969investigating} on the time-series data to detect directional influence among log clusters. The resulting causal graph contains nodes representing log clusters and directed edges denoting inferred causal relationships (Figure~\ref{fig:causal-graph-view}). 

\subsubsection{Tool Usage Workflow}

An effective workflow of our tool begins with the Summary Report, which offers a tabular view of representative log lines annotated with golden signals, fault categories, and entities.
The Temporal Trend guides in identifying the period of fault manifestation, followed by Causal Graph to identify cause-effect relationship, which prioritizes log lines for investigation. Finally, the Diagnosis Report presents relevant log windows within the selected time range for in-depth analysis. This workflow streamlines issue diagnosis, reducing information overload and improving overall productivity.
\section{Experiments}

\begin{figure*}
    \centering
    \begin{subfigure}[t]{0.32\linewidth}
        \includegraphics[width=\linewidth]{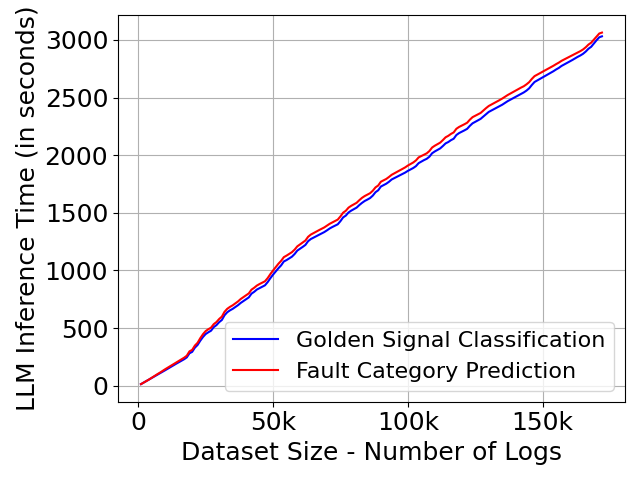}
        \caption{Tool w/o Label Broadcasting}
        \label{fig:logan-without-lb}
    \end{subfigure}
    \hfill
    \begin{subfigure}[t]{0.32\linewidth}
        \includegraphics[width=\linewidth]{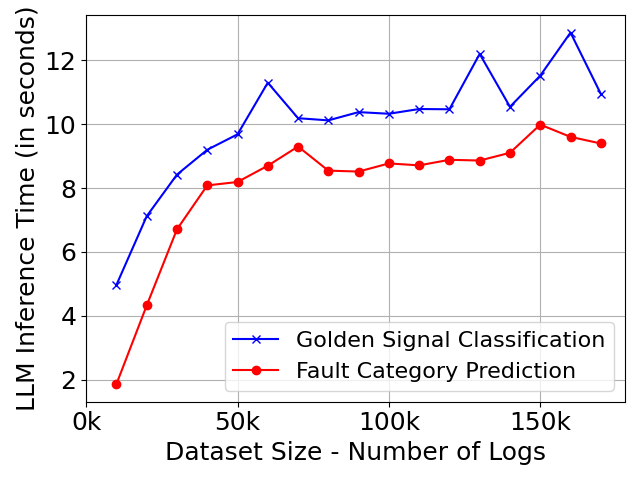}
        \caption{Tool w/ Label Broadcasting}
        \label{fig:logan-with-lb}
    \end{subfigure}
    \hfill
    \begin{subfigure}[t]{0.32\linewidth}
        \includegraphics[width=\linewidth]{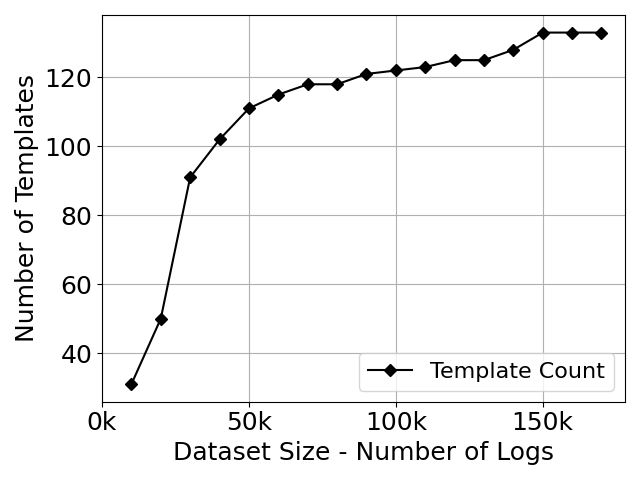}
        \caption{Dataset vs Representation Set Size}
        \label{fig:template-count-vs-size}
    \end{subfigure}

    \caption{Label Broadcasting: Computation Time Analysis}
    \label{fig:rq2-resource-consumption}
\end{figure*}

Our log analytics tool introduces two contributions:
(1)~\textbf{Templatisation with Label Broadcasting (LB):} An efficient and scalable LLM inference technique that runs entirely on CPUs and enriches logs with predicted labels from three tasks - Golden Signal Classification, Fault Category Prediction and Named Entity Recognition, and 
(2)~\textbf{Report Generation:} which leverages the enriched logs to produce reports that reduces the cognitive load on SEs. 

To assess the effectiveness of LB, we address the following research questions:
\textbf{RQ1:} How effective is LB in reducing computation time during LLM Inference? 
\textbf{RQ2:} Does LB affect the LLM Inference Quality?  
To assess the practical utility of the reports generated by our log analytics tool, we present findings from a real case study.

\subsubsection{Dataset Overview}
We evaluate our approach on logs from four real-world software applications (D1, D2, D3, D4) spanning the domains of data analytics, security, inventory management, and finance. More details on the datasets and statistics are provided in Appendix A.3.

As a baseline, we use the traditional method that performs LLM inference on each log line, referred to as \textbf{Tool w/o LB} and compare it with our approach (\textbf{Tool w/ LB}).

\subsubsection{LLM Selection and Rationale}
BERTOps~\cite{gupta2023learning} is an LLM pretrained on logs that has shown strong performance on log analysis tasks in few-shot learning setups. In this study, we adopt BERTOps and evaluate it on the above three tasks. Our results validate that it outperforms competing methods, generalizes well when trained on examples from multiple domains, and remains robust when applied to logs from unseen domains. These findings justify its integration into our tool (refer Appendix A.5). 

\subsection{Results and Discussion}

\subsubsection{RQ1:} 
To analyse our log analytics tool's efficiency, we analyse the computation time required to process log datasets of varying sizes using a Virtual Machine (VM) with $48$ cores and $189$GB RAM, without GPUs. Table~\ref{tab:rq2-time-stats} reports the computation time of our log analytics tool to process a dataset of $170K$ log lines from the D3 Domain (refer Appendix A.3 for details).  
Our approach, Tool w/ LB yields a significant reduction of $\sim99.7$\% in inference time compared to the traditional approach of Tool w/o LB.

\begin{table}[h]
    \small
    \centering
    \begin{tabular}{|c|c|c|c|c|}
    \hline
    \multirow{2}{*}{Method} & \multirow{2}{*}{\thead{\#Data\\Size}} & \multicolumn{3}{c|}{Time Taken (in seconds)} \\ \cline{3-5} 
     &  & \multicolumn{1}{c|}{\thead{Templati-\\zation}} & \multicolumn{1}{c|}{\thead{GSC}} & \thead{FCP} \\ \hline
    \thead{Tool w/o LB} & \multirow{2}{*}{170k} & \multirow{2}{*}{31.1s} & \multicolumn{1}{c|}{3031.1s} & 3063.6s \\ \cline{1-1} \cline{4-5} 
    \thead{Tool w/ LB} &  &  & \multicolumn{1}{c|}{\textbf{10.95s}} & \textbf{9.39s} \\ \hline
    \end{tabular}
    \caption{Influence of Label Broadcasting (LB)}
    \label{tab:rq2-time-stats}
\end{table}

Figure~\ref{fig:logan-without-lb} and ~\ref{fig:logan-with-lb} illustrate the computation time for inferring Golden Signals (GS) and Fault Category (FC) labels as the dataset size increases in increments of $10K$ loglines.
As depicted in Figure~\ref{fig:logan-without-lb}, the computation time for Tool w/o LB increases linearly, which is expected since it infers label for each individual log line.
In contrast, Figure~\ref{fig:logan-with-lb} shows that Tool w/ LB maintains significantly lower processing times for GS and FC label computation, with a maximum of only $10$ seconds. 
Initially, for Tool w/ LB, computation time increases as the dataset grows. 
This is because for each new template discovery by the log templatizer, GS and FC labels are predicted by invoking the LLM, resulting in a steep increase in computation time.
However, after processing approximately $50K$ log lines, the computation time plateaus. 
This is because fewer new templates are identified beyond this point (as shown in Figure~\ref{fig:template-count-vs-size}), allowing our the tool to retrieve GS and FC labels via a simple lookup rather than invoking the LLM model. 
These results indicate that Tool w/ LB achieves substantial resource savings during LLM inference, empirically validating its ability to scale efficiently and handle large-scale log data with minimal increase in computation time.

\begin{figure}[h]
    \centering
    \begin{subfigure}[b]{0.5\textwidth}
        \centering
        \includegraphics[scale=0.44]{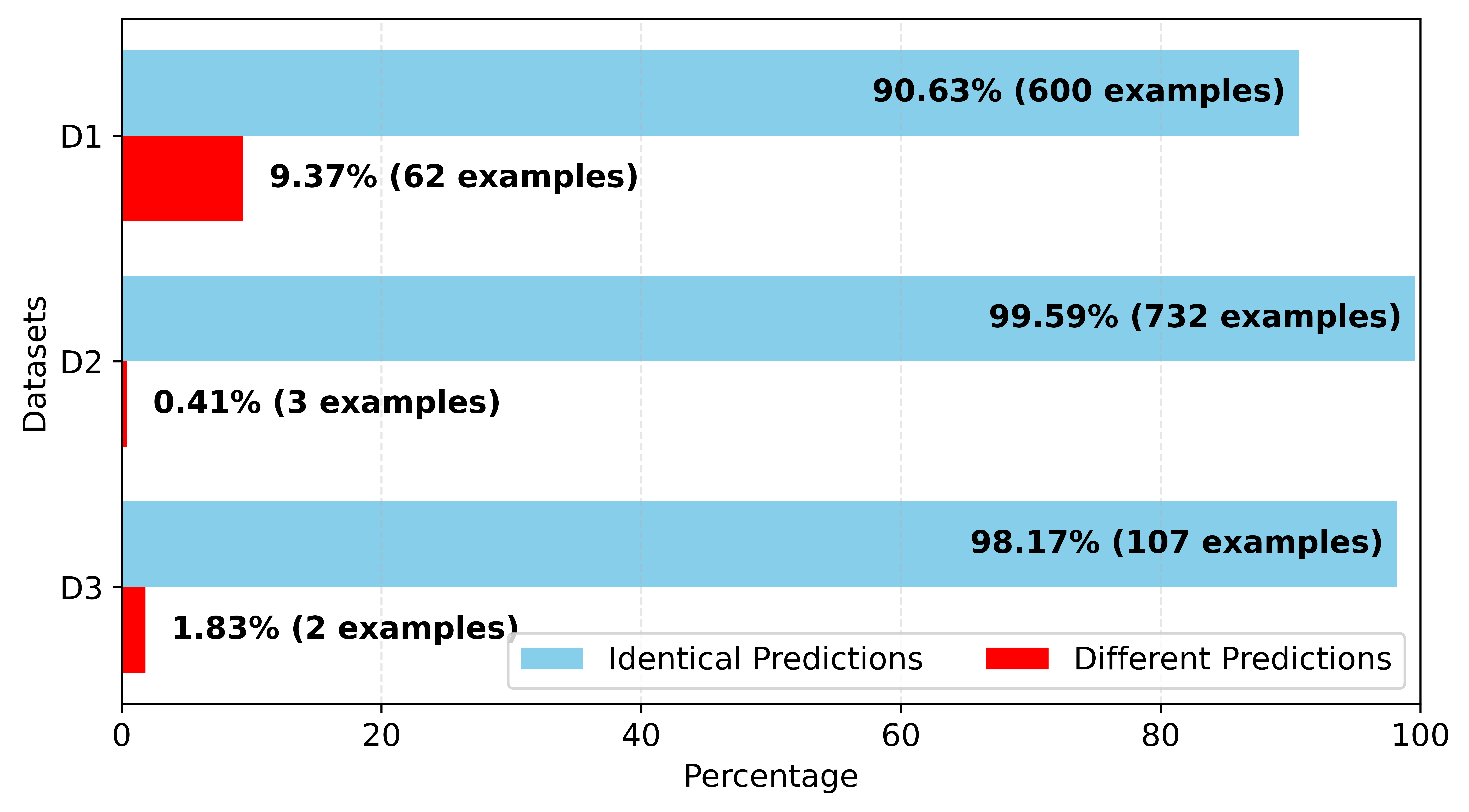}
        \caption{Statistics: Identical vs Different Predictions}
        \label{fig:rq1-barchart-overall}
    \end{subfigure}
    \begin{subfigure}[b]{0.5\textwidth}
        \centering
        \includegraphics[scale=0.55]{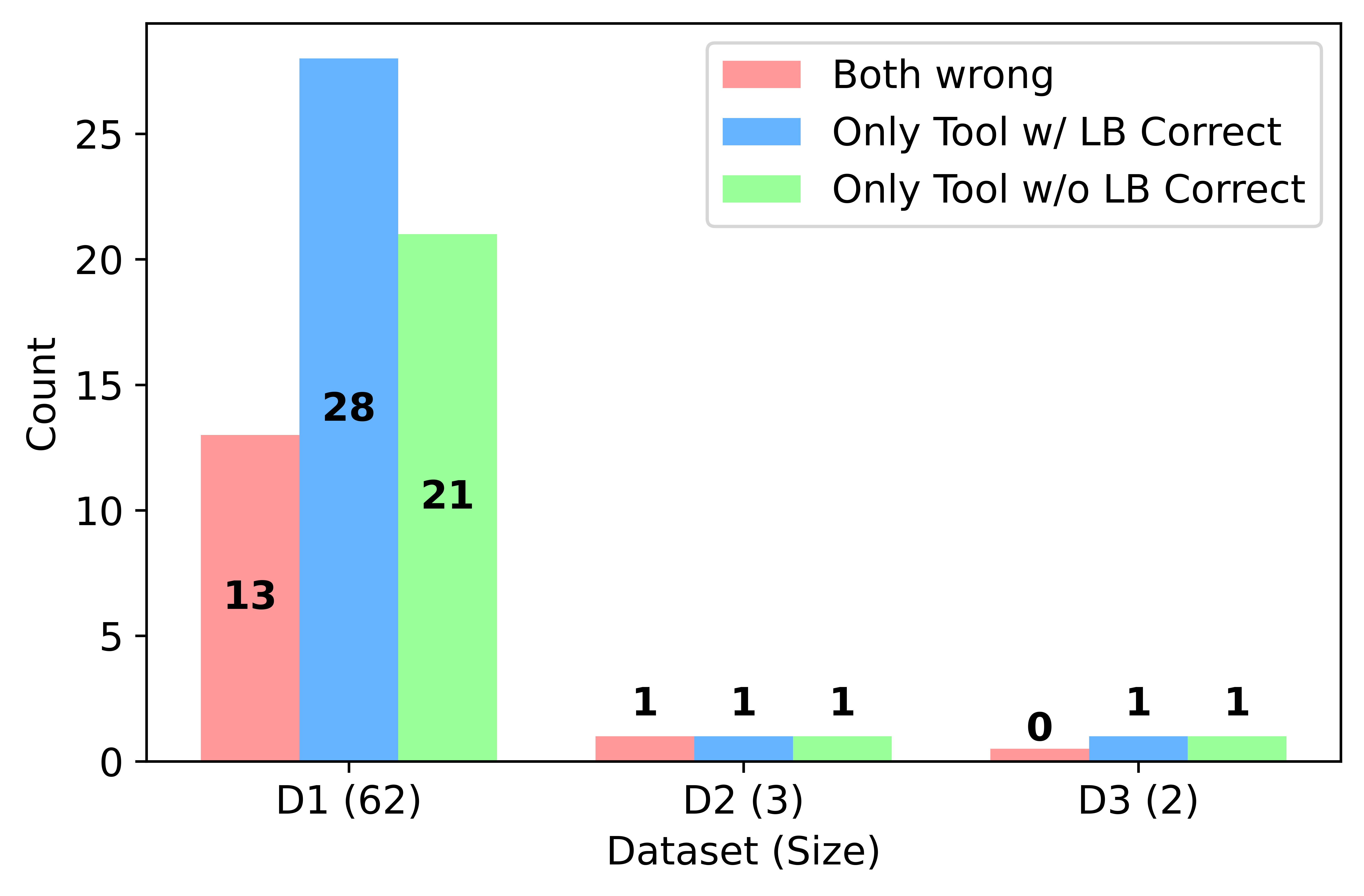}
        \caption{Different Prediction Cases: Tool w/ LB \& Tool w/o LB}
        \label{fig:rq1-barchart-differ}
    \end{subfigure}

    \caption{Label Broadcasting: Output Comparison Analysis}
    \label{fig:rq1-comparison}
\end{figure}




\subsubsection{RQ2:} 
To evaluate the effect of Label Broadcasting (LB) on LLM inference quality, we compare the predictions generated by the two methods - Tool w/ LB and Tool w/o LB - across the three datasets (D1, D2, and D3). Figure~\ref{fig:rq1-barchart-overall} presents the comparison results.
Across all the datasets, the two methods produce identical outputs for the majority of the cases. For datasets D2 and D3, over $98\%$ of log lines yield the same inference output under both methods. This indicates that applying Tool w/ LB doesn't alter the outcome compared to Tool w/o LB, thereby leaving LLM inference quality unaffected for these cases.
The figure also highlights the small proportion of cases where Tool w/ LB and Tool w/o LB produce different predictions. For D2 and D3, such instances are extremely rare, containing only $3/735$ and $2/109$ examples, respectively. In contrast, dataset D1 is an exception, with predictions differing in $9.37\%$ of the cases.

Figure~\ref{fig:rq1-barchart-differ} further analyses these differing cases in D1 by comparing each method's predictions against the ground truth, resulting in three categories:
(1) \textbf{Both wrong:} In $2\%$ of cases ($13$ examples), both methods produce incorrect predictions, indicating that the model fails to generate the correct label regardless of the approach. Here, Tool w/ LB is preferable as it reduces the number of LLM invocation calls.
(2) \textbf{Only Tool w/ LB correct:} In $4.2\%$ of cases ($28$ examples), Tool w/ LB predicts the correct label, while Tool w/o LB is incorrect. In these cases, Tool w/ LB not only improves LLM inference quality but also reduces resource usage by minimizing LLM calls.
(3) \textbf{Only Tool w/o LB correct:} In $3.2\%$ of cases ($21$ examples), Tool w/o LB predicts the correct label while Tool w/ LB is incorrect. Examples from this category are shown in Table 9 (refer Appendix), suggesting that variables in log lines offer \textit{important} cues that help the LLM make accurate predictions.
For such rare cases, performing inference on each log line individually is advantageous, as Tool w/ LB can negatively impact prediction quality. Nevertheless, in the remaining $96.8\%$ of cases, Tool w/ LB yielded the \textit{preferred} outputs.

As supported by our experiments and prior work~\cite{logmid}, the occurrence of such \textit{important} variables is rare. 
Since Tool w/ LB produces \textit{preferred} predictions for nearly $96\%$ of log lines, the relatively rare cases in category 3 are considered an acceptable operational trade-off. Achieving perfect accuracy for these few instances would require GPU-based inference on every log line, which would significantly increase resource requirements.
In contrast, Tool w/ LB enables our log analytics tool to operate entirely on CPU, greatly reducing its memory footprint while scaling efficiently to handle large datasets across multiple products. Hence, our experimental results validate that Label Broadcasting can be employed for LLM inference for logs without degrading overall prediction quality.

\subsection{Case Study}
To assess the practical utility of our log analytics tool, we present a real-world case study from an application in the financial domain (D4). The client provided log data from July 3–5, 2023, consisting of $425{,}000$ log lines.
The client reported that the application was encountering issues, resulting in unexpected terminations, but they were unclear how to go about diagnosing it. Their request made to us was: 
\textit{``We need a way to correlate and make sense of this data, at least to reduce the cognitive load on our subject matter experts (SMEs) and provide actionable insights. 
We need conclusive, data-driven evidence for such correlations."}
The following discussion highlights the effectiveness of our tool in debugging and diagnosing issues.

\begin{figure*}[!htp]
    \centering
    \includegraphics[width=\textwidth, trim={1cm 0 1cm 0},clip]{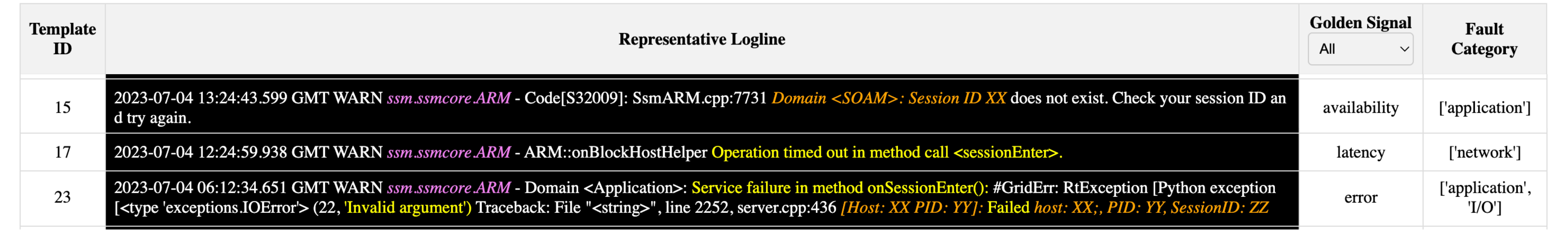}
    \caption{Summary Report showing representative log lines along with their golden signal, fault categories and named entities. }
    \label{fig:summary-view}
\end{figure*}

\textbf{(a) Summary Report:} 
Our tool reduced the dataset from $425{,}000$ log lines to a representative set of just $74$ log lines, achieving a reduction of $99.9\%$. The golden signals distribution within this set was: \textit{\{error:18, availability:9, latency:5, saturation:2, information: 40\}}. Since the log lines under \textit{information} signal mainly contains debugging context, these can be safely ignored~\cite{gupta2023learning} for initial issue diagnosis. Among the remaining log lines, the high count of error and availability signals highlight them as prominent concerns. 
Figure~\ref{fig:summary-view} shows a snapshot of the summary report with their color-coded entities, golden signal and fault categories. 
To illustrate its utility, we detail the log lines by explicitly indicating which log analysis task extracted the corresponding information (shown in brackets). For instance, Template ID (TID) 15: ``Session ID XX does not exist" (NER) correspond to an ``availability" issue (GSC) at the ``application" level (FCP). Similarly, for TID 17: ``Operation Timed out" (NER) resulting in ``latency" issue (GSC) at the ``network" stack (FCP).

Moreover, our tool also extracted critical attributes such as host/application name and process/session identifiers. For example, TID 23, ``Service Failure" caused by ``invalid argument" for ``Host XX, PID YY, SessionID ZZ" (NER) is categorized as an ``error" (GSC) at the ``application and I/O" level (FCP). This significantly reduces the cognitive load on SEs by highlighting key entities that would otherwise require manual search, thus improving productivity.

\begin{figure}[!htp]
    \centering
    \includegraphics[width=0.475\textwidth]{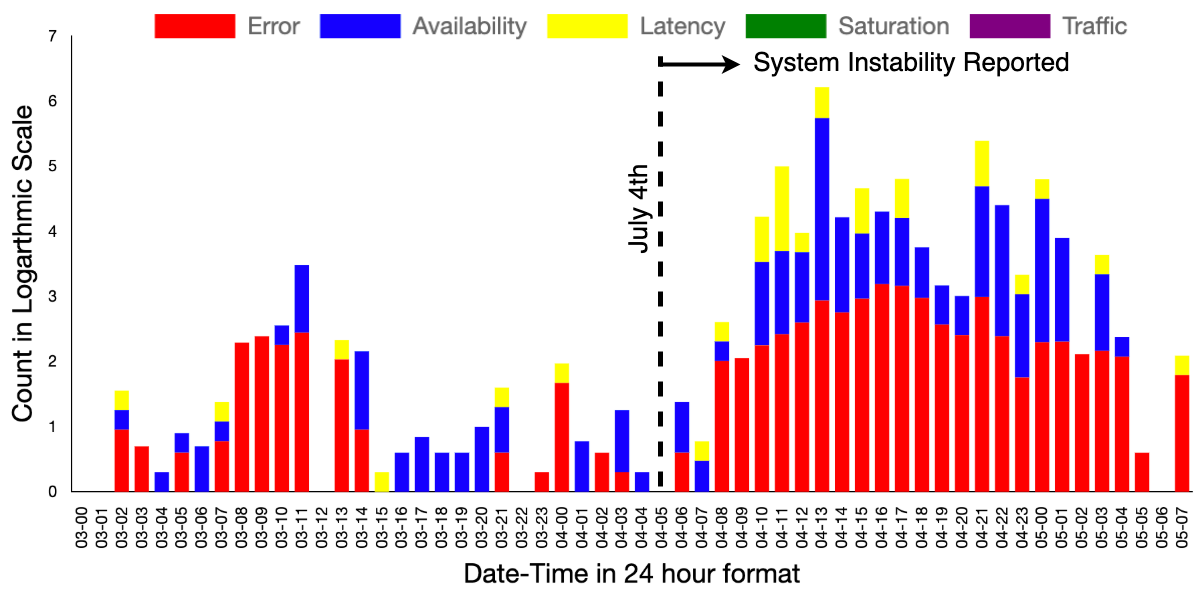}
    \caption{Temporal View of Fault Progression} 
    \label{fig:temporal-trend-view}
\end{figure}

\textbf{(b) Temporal Trend View}: Figure~\ref{fig:temporal-trend-view} presents the hourly temporal trend of various golden signals in the log data, tracking issue progression and its manifestation. For instance, the trend indicates a sharp escalation after July 4th (vertical black line), when the client system's instability became more pronounced. Before July 4th, there were only a few errors and availability events. However, these early signs of trouble gradually intensified after July 4th, leading to increased occurrences of errors, availability and latency events, signalling a broader system degradation. 

\begin{figure}[h]
    \centering
    \includegraphics[width=0.47\textwidth, trim={0.5cm 0 0.5cm 0},clip]{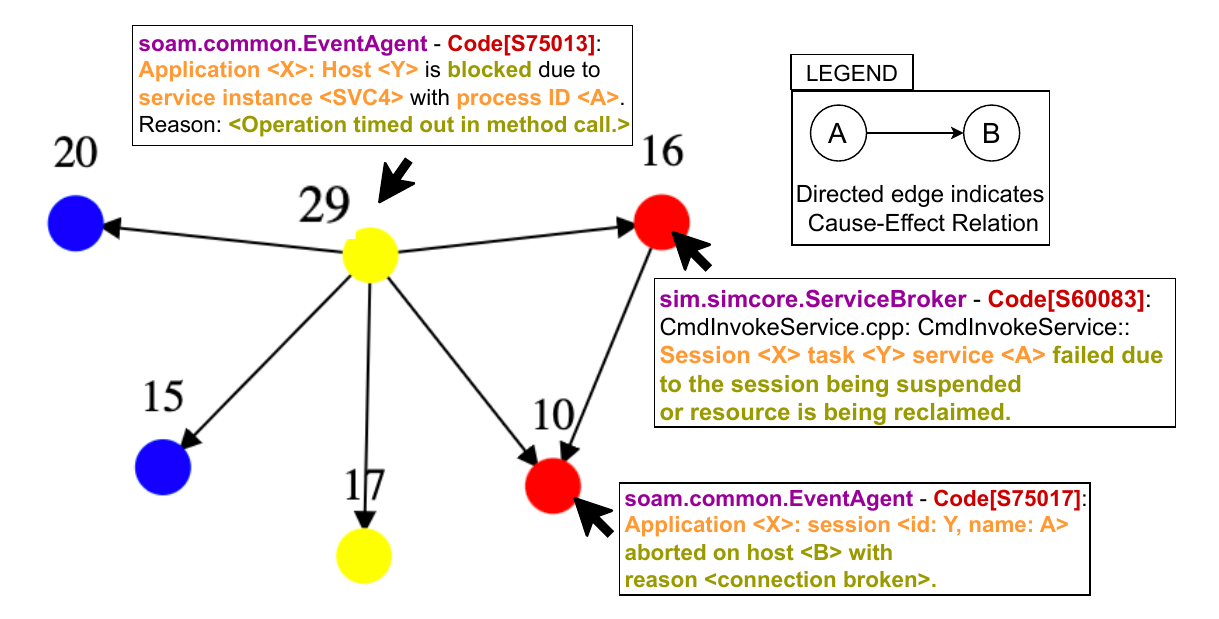}
    \caption{Causal Graph View}
    \label{fig:causal-graph-view}
\end{figure}

\textbf{(c) Causal Graph View}: Figure~\ref{fig:causal-graph-view} depicts the causal graph view generated by the tool, where the nodes follow the same colour scheme as the Temporal Trend View (Figure~\ref{fig:temporal-trend-view}). 
In the UI, hovering over a node reveals the corresponding representative log line, with its color-coded entities.
From the graph, it is evident that the tool identified cause–effect relations among six nodes. A closer inspection highlights the following insights: (1) TID 29: “Operation Timed Out” (NER) associated with ``latency" issue (GSC).
(2) TID 10: “Connection Broken” (NER) associated with “error” (GSC), and  (3) TID 16: “Session being suspended” (NER) associated with  ``error" (GSC). 
A plausible explanation for this correlation is that latency observed in TID 29 leads to suspended session in TID 16, which eventually results in the broken connection error in TID 10.
The graph thus reveals connections among logs that might otherwise appear unrelated.
This capability enables SEs to identify hidden patterns for more effective issue diagnosis.



In essence, the Summary Report presented diverse insights extracted via LLM inference while reducing the log data volume by $99.9\%$. The Temporal Trend indicates that system degradation began around July 4th, whereas the Causal Graph highlights latency as a prevalent concern. Building on these insights, SEs can leverage the Diagnosis Report (Figure 11 in Appendix D) to focus only on the \textit{relevant} log windows and conduct deeper analysis on the \textit{problematic} log lines identified by the other reports. These findings, along with explanations and reasoning, were shared with the client, who appreciated the tool's contributions for effective issue diagnosis.

\section{Lessons Learned From Deployment}

This section outlines the system design of our log analytics tool’s deployment pipeline and shares key insights and lessons learned from its adoption across $70$ IBM Software Support products.


\begin{figure}[h]
    \centering
    \small
    \includegraphics[width=0.47\textwidth, trim={0.5cm 0.21cm 0.5cm 0.21cm},clip]{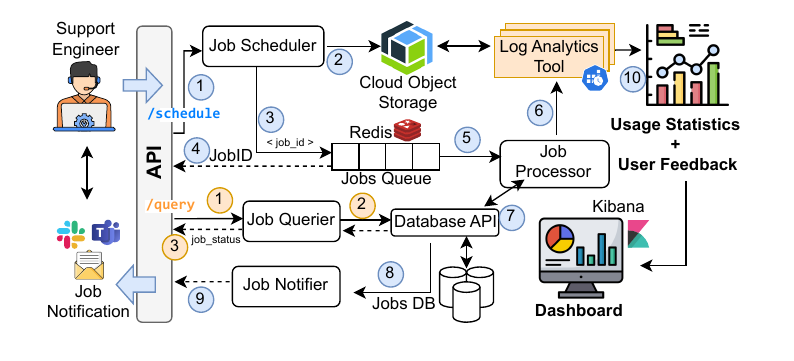}
    \caption{System Design: Workflow of \texttt{/schedule} API (blue, 1–10) and \texttt{/query} API (orange, 1–3) }
    \label{fig:sys-design}
\end{figure}

\begin{figure*}
    \small
    \centering
    \begin{subfigure}[t]{0.32\textwidth}
        \centering
        \includegraphics[width=\textwidth]{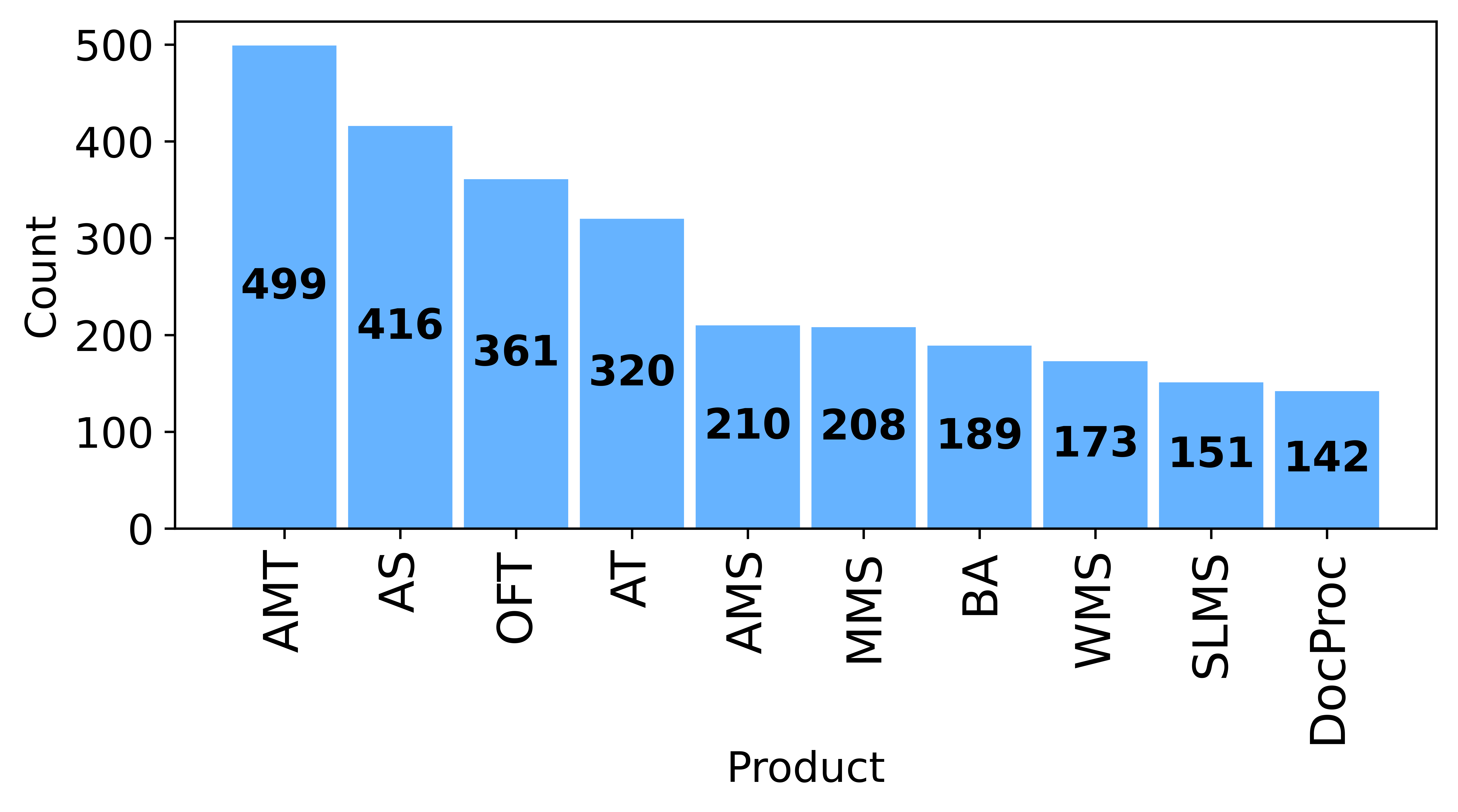}
        \caption{Top 10 Products Based on Usage}
        \label{fig:logan-top10}
    \end{subfigure}
    \hfill
    \begin{subfigure}[t]{0.32\textwidth}
        \centering
        \includegraphics[width=\textwidth]{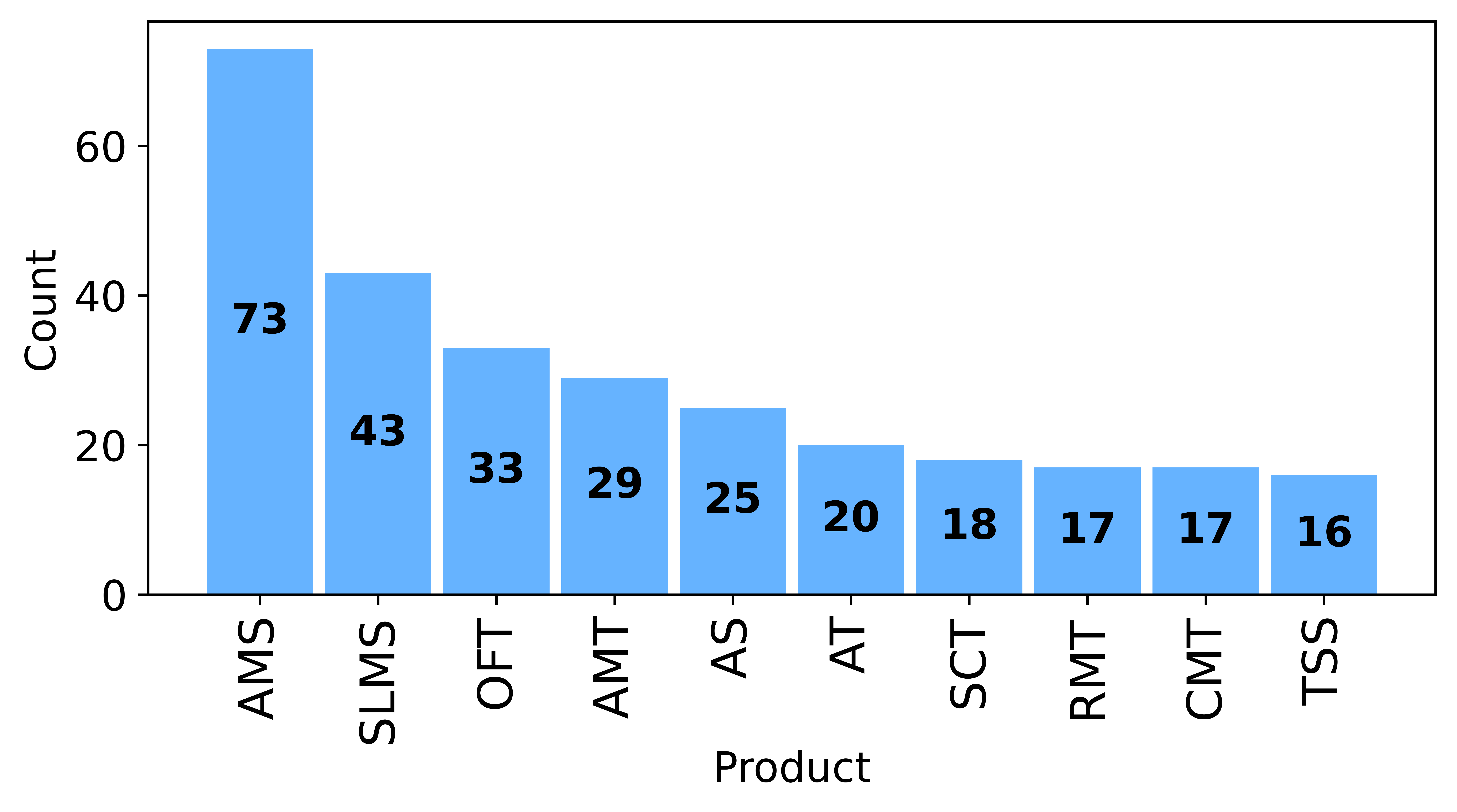}
        \caption{Top 10 Products Benefiting Most from the Tool}
        \label{fig:logan-useful}
    \end{subfigure}
    \hfill
    \begin{subfigure}[t]{0.32\textwidth}
        \centering
        \includegraphics[width=\textwidth]{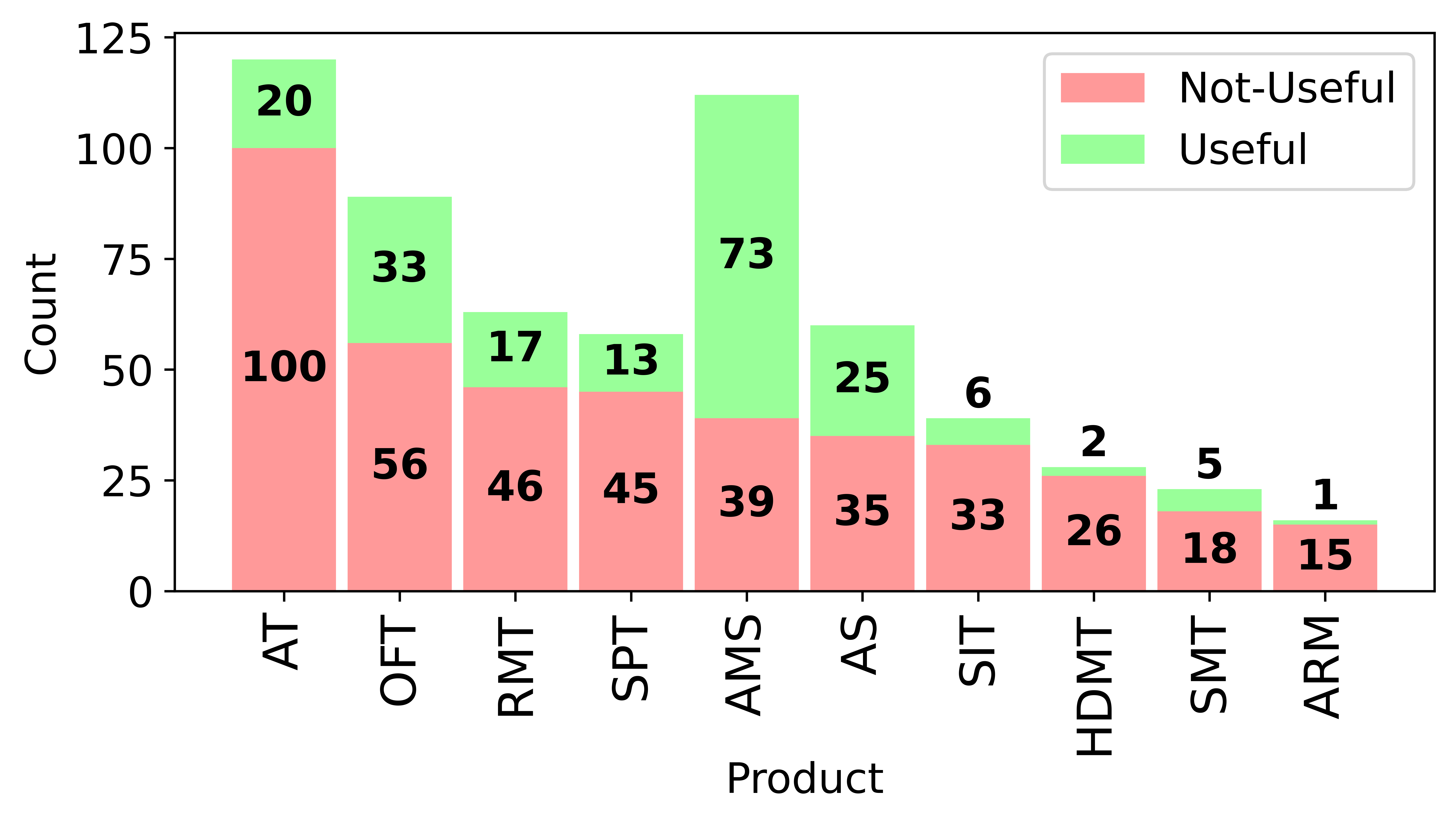}
        \caption{Top 10 Products that didn't find the Tool useful}
        \label{fig:logan-notuseful}
    \end{subfigure}
    \label{fig:logan-overall}
    \caption{\textbf{Summary of Product Usage with the Log Analytics Tool.} 
(Acronyms\footnotemark: 
AMS: Application Monitoring Solution, 
AMT: Asset Management Tool, 
ARM: Application Resource Management, 
AS: Application Server, 
AT: Automation Tool, 
BA: Business Automation, 
CMT: Content Management Tool, 
DocProc: Document Processing Software, 
HDMT: Device Management Tool, 
MMS: Microservice Management Tool, 
OFT: Order Fulfillment Tool, 
RMT: Requirements Management Tool, 
SCT: Software Collaboration Tool, 
SIT: Software Integration Tool, 
SPT: System Performance Tool, 
SLMS: Software Lifecycle Management Solution, 
SMT: System Monitoring Tool, 
TSS: Test Suite Software, 
WMS: Workplace Management System.)}
\end{figure*}

\subsection{System Design}

Figure~\ref{fig:sys-design} depicts the microservices-based architecture of our log analytics tool's pipeline.
A SE uploads a log dump (hereby known as \textit{job}), via the \texttt{/schedule} API endpoint. 
The Job Scheduler assigns a unique \textit{job\_id} to each uploaded log dump and stores the data in Cloud Object Storage (COS). Then the \textit{job\_id} is placed in the Job Queue, and the user is notified that the job has been scheduled. When the job is ready for execution, the Job Processor retrieves the job ID, launches an independent instance of our log analytics tool, and begins processing. Multiple jobs are processed concurrently, with their statuses tracked in a database (Jobs DB). Users receive status updates via configured channels such as Slack or Microsoft Teams, or by querying the \texttt{/query} endpoint. Additionally, runtime and usage statistics—such as total processing time, number of files, and data size—are stored in Elasticsearch and visualized through a Kibana dashboard, providing stakeholders with insights for monitoring and iterative improvement.

\footnotetext{Product names are obfuscated.}
\subsection{Lessons Learned - Deployment}

Since March 2024, the log analytics tool has been deployed in production, with $70$ products onboarded to leverage it for analyzing customer tickets containing log data.
\begin{figure}[h]
    \small
    \centering
    \includegraphics[width=0.5\linewidth]{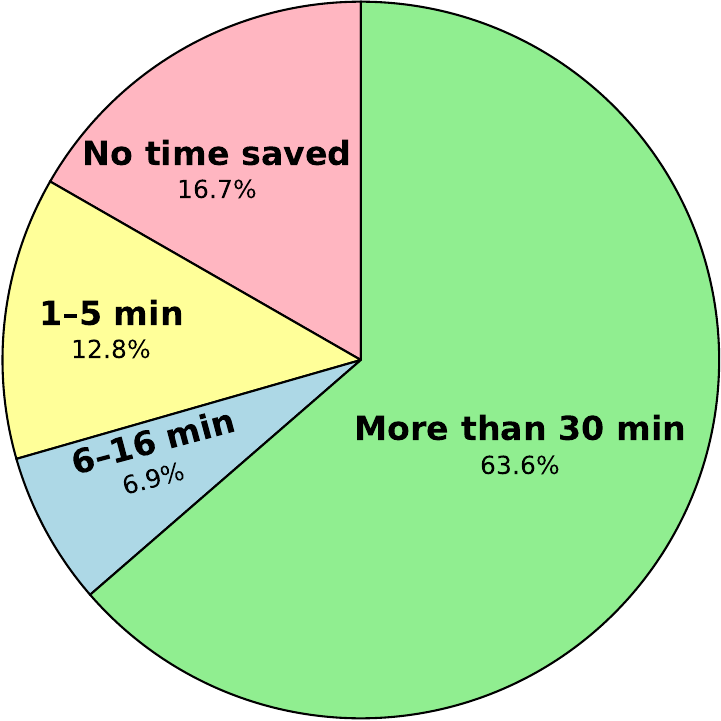}
    \caption{Distribution of time savings achieved with our tool.}
    \label{fig:time-saved-distribution}
\end{figure}
Figure~\ref{fig:logan-top10} highlights the top $10$ products that used the  tool for log analysis. The number of times the tool was triggered per product ranges from a minimum of $1$ to a maximum of $499$ times, with an average of $69$ times. Over the $15$ months, the tool has been invoked $4,227$ times across $2,394$ unique tickets, processing approximately $1.49$ TB of data and $2.76$ billion log lines. On average, each ticket involved a $653$ MB log dump and approximately $1$ million log lines. The tool has spent an average of $50$ minutes processing each ticket end-to-end (refer Production Usage Dashboard in Appendix B).


To assess the time savings from using the tool, we solicited feedback from SEs using a brief three-question survey at the end of each session:
\begin{itemize}
\item Q1. Was the final output useful? (Yes/No)
\item Q2. How much time did the tool save? a) No Savings, b) $1-5$ minutes, c) $6-15$ minutes, d) More than $30$ minutes
\item Q3. Open-ended feedback: What additional features or improvements would make the tool more useful?
\end{itemize}
The first question assesses whether the tool aided SEs in issue diagnoses. The second evaluates whether the tool contributed to time savings. The third question elicit user feedback to identify shortcomings and gather feature requests. This structured feedback approach enabled us to measure the tool's impact and identify opportunities for improvement.

As the feedback was optional, we received responses for $565$ tickets, representing $23.51\%$ of the total tickets. Of these responses, $46.72\%$ of the respondents found the tool useful. Some of the positive and negative feedbacks along with feature requests are illustrated in Appendix C. The top $10$ products that received positive feedback are illustrated in Figure~\ref{fig:logan-useful}. These products resulted in savings of $316$ man hours, with the distribution in terms of savings illustrated in Figure~\ref{fig:time-saved-distribution}. A major portion is attributed to savings of more than 30 minutes. 

Among the $53.24\%$ of the population that did not find the tool useful, the respective top $10$ product distribution is illustrated in Figure~\ref{fig:logan-notuseful}. The reason for equally high negative sentiment can be attributed to negativity bias. Individuals are generally more likely to provide feedback after negative experiences than positive ones, a tendency explained by negativity bias \cite{baumeister2001bad,anderson1998customer,oliver1980cognitive}, where adverse events exert greater cognitive and behavioral impact. As a result, feedback may overrepresent negative sentiments, potentially skewing perceptions of a tool’s effectiveness. In this case, however, the gap between negative and positive feedback is small, suggesting that many users valued and found the tool useful.


To further analyze negative feedback, we identified the top $10$ products where users found the tool not useful. For these products, we also calculated how often it was marked useful. Figure~\ref{fig:logan-notuseful} shows that in $9$ of $10$ cases, “not useful” counts exceed “useful,” indicating consistent usability or applicability issues. This pattern indicates a possible misalignment between the tool’s capabilities and the logging patterns or diagnostic needs of these products. AMS is a positive outlier, as more people find it useful than not. For all the products illustrated in Figure~\ref{fig:logan-notuseful}, we manually reviewed some of the tickets for each product and examined their log dumps. Also, we interviewed the SEs who worked on those tickets and based on their responses to Q3, the key reasons are: 
\begin{enumerate}
    \item Some users, especially those who are domain experts, believe that when provided with the log data and problem description, they are better equipped to manually filter related files to diagnose the underlying issue.
    \item Expert users felt that the tool processing was excessive; given their expertise in handling such cases, they could manually diagnose the issue in the same amount of time.
    \item We found that for three products - Software Integration Tool (SIT), Application Resource Management (ARM), and Order Fulfilment Tool (OFT) - the log dumps were significantly large, ranging from several tens of gigabytes. As these are microservice-based applications, they generate multiple files, one for each pod and service. Processing such a large volume of files demands considerable time and resources, slowing the overall log analytics pipeline and, in some cases, causing it to crash entirely.
    \item Manual investigation of log dumps provided further insights into data quality. For the product Hardware Device Management Tool (HDMT), the logs were mostly related to hardware (disk), very different from the kind of data that the models were trained on. The log analytics tool is appropriate for software application logs and not hardware logs. Similarly, logs from mainframe-based applications were not very effective for the tool to process.
    \item Another aspect of the quality of log data that affected the efficacy of the log analytics tool is related to the presence of JSON objects embedded with textual data. Such kind of intermingling of JSON objects with textual data jeopardises the templatization algorithm, leading to too many clusters, and gains from label broadcasting are not efficient, thereby slowing the entire pipeline.
\end{enumerate}

In summary, over a $15$-month deployment in production, the tool has scaled to $70$ products, efficiently processing $2,394$ tickets, with an average ticket size of $700$ MB, resulting in savings of $316$ man-hours. Our novel approach demonstrates the efficiency and effectiveness of processing large volumes of log data with LLMs in a CPU-constrained environment. Usage patterns suggest two main future directions: (a) processing logs more efficiently, and (b) better handling of JSONs embedded in logs to improve log templates quality, and processing speed.

\section{Related Works}

This section presents the related work in two directions: 

\textbf{(i) Use of LLMs in log-related tasks: }
Recent works have used LLMs to perform various log-related tasks. Some studies used LLMs for log anomaly detection~\cite{logfit, lad, LAD_24, LAD_1_2024}. Almodovar et al. continually pretrain RoBERTa with log data for anomaly detection~\cite{logfit}. Liu et al. conducted a case study on logs and found that the anomalies detected by ChatGPT were partially aligned with those identified by on-call engineers ~\cite{lad}. Other studies have leveraged LLMs for log parsing~\cite{llmparser, logparser1, divlog, parser, lilac, librelog, lunar}. Ma et al. proposed an LLM-based log parser based on generative LLMs and few-shot tuning~\cite{llmparser}. Zhong et al. leveraged LLMs for log parsing, merging syntactic and semantic insights~\cite{logparser1}. 
Most of these papers utilise decoder models, and none examine case studies on how such models are deployed in real-world production environments with employing CPUs for inferencing. Most recently, Liu et. al ~\cite{LogLM:FromTaskBased} introduced an instruction-tuned model for diverse log analysis tasks into a single, unified instruction-response format. The fine-tuned model on LLaMA2-7B is deployed within a software and network platform of Huawei. However, no further analysis on generalizability, user feedback, time complexity or productivity has been reported. 

\textbf{(ii) Enterprise-grade frameworks that provide log analysis and discovery capabilities:} Other works ~\cite{metalearning, logai} and several prominent products in the community (including New Relic~\cite{newrelic}, Dynatrace~\cite{dynatrace}, DataDog~\cite{datadog}, Instana~\cite{instana} offer comprehensive log monitoring platforms and frameworks.  However, none of these works currently integrate LLMs into their frameworks at scale. Also, while these tools share similar functionalities, we focus on a detailed comparison with one well-known product, New Relic, to illustrate the distinct advantages of our log analytics tool. NewRelic displays log streams generated by applications, highlights basic entities in the log lines such as service names and log levels, shows the temporal evolution of logs, and detects anomalies but doesn’t explain the cause and has no human-readable summaries generated. The proposed log anlaytics tool, on the other hand, displays relevant \textit{windows} of correlated log streams (Diagnosis View); highlights complex entity types such as cause/symptom phrases, process ID, component, etc. (through Summary View); shows a more comprehensive temporal evolution which includes insights from the golden signal, and fault categories, correlates logs from different microservices to derive causal relations among log lines which helps in effective and efficient issue diagnosis.

\section{Conclusions and Future Work}
This paper presents a log analytics tool that uses a Large Language Model for log analysis and causal graph mining from log data. One of the defining characteristics of the tool is that it can use LLM to process log data originating from disparate software applications, in time defined manner, with minimum resource consumption - running on CPU, and yet not compromising on the final output. Over 15 months in production, our tool scaled across 70 products, processing 2,394 tickets (avg. 700 MB), saving 316 man-hours. Results show the effectiveness of our LLM-based approach in CPU-constrained environments, with future work focusing on more efficient and faster way of preprocessing logs.


\bibliography{references.bib}

\clearpage
\appendix
\section{Supplementary Material}

\section{A. LLM Integeration and Rationale}

In this section, we present the rationale for integrating BERTOps~\cite{gupta2023learning} LLM into our log analytics tool. We begin by describing the three log analysis tasks used by our tool -  Golden Signal Classification (GSC), Fault Category Prediction (FCP) and Named Entity Recognition (NER).We then outline the experimental design of our study, followed by details of the task-specific datasets curated and the baseline models used for comparison. Finally, we report the experimental results and discuss their key findings.

\subsection{A.1 Log Analysis Tasks}
This section defines the three tasks that our log analytics tool leverages to extract insights. For each task, we fine-tune a dedicated LLM, resulting in three task-specific models that are subsequently integrated into the tool.

\subsubsection{Golden Signal Classification (GSC)}
The Google Site Reliability Engineers (SRE) handbook~\cite{beyer2016site} defines six ``Golden Signals" for monitoring cloud applications - Latency, Traffic, Error, Saturation, Availability and Information. 
The first five serve as primary indicators for detecting \textbf{problematic} log lines, while the \textit{information} signal provides debugging insights.

\subsubsection{Fault Category Prediction (FCP)}
Fault categories in logs act as signals for detecting anomalies, aiding failure diagnosis, and routing issues to relevant teams. A log line may exhibit one or more of seven categories: memory, network, authentication, I/O, device, application, and other~\cite{zou2014improving}.

\subsubsection{Named Entity Recognition (NER)}
Identifying key entities in log lines enable support engineer focus at critical entities, thereby facilitating structured analysis. Table~\ref{tab:ner-example} presents an exemplar log line with entities extracted by our tool. 

\begin{table}[!htp]
\small
\centering
\begin{tabular}{|cc|}
\hline
\multicolumn{1}{|c|}{\textbf{Entity Type}} & \textbf{Parsed Text}\\ \hline
\multicolumn{1}{|c|}{DateTime} & 20/10/24 14:30 \\ \hline
\multicolumn{1}{|c|}{Level} & ERROR \\ \hline
\multicolumn{1}{|c|}{ProcessID} & 123 \\ \hline
\multicolumn{1}{|c|}{ErrorCode} & 500 \\ \hline
\multicolumn{1}{|c|}{URL} & https://ABC.com/api/data \\ \hline
\multicolumn{1}{|c|}{Cause} & Connection failure \\ \hline
\multicolumn{1}{|c|}{Symptom} & Unable to fetch data \\ \hline
\multicolumn{1}{|c|}{NV-Pair} & userID=6789. \\ \hline
\multicolumn{1}{|c|}{FileOrDir} & /db.json \\ \hline
\multicolumn{2}{|p{8cm}|}{\raggedright \textbf{LogLine:} 2024-10-20 14:30 ERROR [ErrorCode: 500] [ThreadID - 123] [Config: /db.json] Unable to fetch data from https://ABC.com/api/data. Connection failure for userID=6789} \\ \hline
\end{tabular}
\caption{Example of entities extracted from logs}
\label{tab:ner-example}
\end{table}


\subsection{A.2 Experimental Design}
The choice of model for our log analytics tool is determined by its ability to \textbf{generalize} on logs from multiple domains and also on its \textbf{robustness} in handling domain shifts from logs generated by diverse sources.

Evaluating model performance on the three tasks requires manually annotated data, which is inherently difficult to obtain. Annotation is both time-consuming and labor-intensive, particularly in industrial settings. Moreover, logs from real incident cases often contain client-specific sensitive information, making it impractical to share them for labeling. To address these challenges, we evaluate the models’ ability to make accurate predictions with minimal labeled data. This motivates our first experiment \textbf{(1)}: How does a model perform in a few-shot setting, where only a small amount of labeled data is available for fine-tuning task-specific models?

As the log analytics tool is applied on application logs from multiple domains, maintaining the accuracy and effectivness of the finetuned models become a critical challenge. 
Since these models are trained on small set of examples in few-shot learning setup, their performance may vary when encountering unfamiliar or previously unseen data from new domains. 
In such cases, model retraining may be required to ensure the tool remains effective. 
This dynamic scenario motivates our second experiment \textbf{(2)}: How robust and generalizable is the model when applied to unseen data?

The rest of the section provides details about the datasets, baseline models, and evaluation criteria.

\begin{table}[]

\centering
\resizebox{.48\textwidth}{!}{%
\begin{tabular}{|p{1cm}|ccccccccc|}
\hline
\multirow{3}{*}{Dataset} & \multicolumn{9}{c|}{Tasks} \\ \cline{2-10} 
 & \multicolumn{3}{c|}{GSC} & \multicolumn{3}{c|}{FCP} & \multicolumn{3}{c|}{NER} \\ \cline{2-10} 
 & \multicolumn{1}{c|}{Tr} & \multicolumn{1}{c|}{V} & \multicolumn{1}{c|}{Te} & \multicolumn{1}{c|}{Tr} & \multicolumn{1}{c|}{V} & \multicolumn{1}{c|}{Te} & \multicolumn{1}{c|}{Tr} & \multicolumn{1}{c|}{V} & Te \\ \hline

\multirow{2}{*}{\textbf{\begin{tabular}[c]{@{}c}\thead{Public\\Data}\end{tabular}}} & \multicolumn{3}{c|}{(GSBench)} & \multicolumn{3}{c|}{(FCBench)} & \multicolumn{3}{c|}{(NA)} \\ 
 & \multicolumn{1}{c|}{150} & \multicolumn{1}{c|}{85} & \multicolumn{1}{c|}{338} & \multicolumn{1}{c|}{210} & \multicolumn{1}{c|}{174} & \multicolumn{1}{c|}{694} & \multicolumn{1}{c|}{-} & \multicolumn{1}{c|}{-} & - \\ \hline

\multicolumn{1}{|c|}{\textbf{D1}} & \multicolumn{1}{c|}{30} & \multicolumn{1}{c|}{164} & \multicolumn{1}{c|}{662} & \multicolumn{1}{c|}{130} & \multicolumn{1}{c|}{158} & \multicolumn{1}{c|}{639} & \multicolumn{1}{c|}{39} & \multicolumn{1}{c|}{557} & 2259 \\ \hline

\multicolumn{1}{|c|}{\textbf{D2}} & \multicolumn{1}{c|}{20} & \multicolumn{1}{c|}{182} & \multicolumn{1}{c|}{735} & \multicolumn{1}{c|}{80} & \multicolumn{1}{c|}{171} & \multicolumn{1}{c|}{695} & \multicolumn{1}{c|}{15} & \multicolumn{1}{c|}{2193} & 8775 \\ \hline

\multicolumn{1}{|c|}{\textbf{D3}} & \multicolumn{1}{c|}{20} & \multicolumn{1}{c|}{26} & \multicolumn{1}{c|}{109} & \multicolumn{1}{c|}{50} & \multicolumn{1}{c|}{25} & \multicolumn{1}{c|}{102} & \multicolumn{1}{c|}{38} & \multicolumn{1}{c|}{39} & 210 \\ \hline

\multicolumn{1}{|c|}{\textbf{D4}} & \multicolumn{1}{c|}{30} & \multicolumn{1}{c|}{17} & \multicolumn{1}{c|}{71} & \multicolumn{1}{c|}{90} & \multicolumn{1}{c|}{14} & \multicolumn{1}{c|}{59} & \multicolumn{1}{c|}{8} & \multicolumn{1}{c|}{11} & 55 \\ \hline
\end{tabular}}
\caption{Dataset Statistics for Golden Signal Classification (GSC), Fault Category Prediction (FCP), Named Entity Recognition (NER) across Train(Tr), Validation(V), Test(Te) Splits}
\label{tab:combined-stats}

\end{table}

\subsection{A.3 Datasets}
\label{sec:datasets}
For evaluation, we collected logs from four real-world software applications to assess model performance across diverse operational environments. Table~\ref{tab:combined-stats} reports the dataset statistics for GSC, FCP and NER tasks under few-shot settings. 
\begin{itemize}
    \item \textbf{Dataset 1} (D$1$) consists of logs collected from a data analytics product that enables organizations to collect, organize, and analyze data across multi-cloud environments.
    \item \textbf{Dataset 2} (D$2$) consists of logs collected from an access management solution that provides organizations with secure authentication, authorization, and access control for their applications.
    \item \textbf{Dataset 3} (D$3$) consists of logs collected from an enterprise asset management solution designed to help organizations manage and optimize the lifecycle of their physical assets. 
    \item \textbf{Dataset 4} (D$4$) consists of logs collected from a financial service product that offers a wide range of services, including wealth management and investment solutions. 
\end{itemize}   
The four products were composed of additional middleware and databases such as MongoDB, Kafka, Db2, Websphere, etc. Consequently, the log data comprised application, middleware, and infrastructure logs. These real-world datasets have few labeled examples per task and uneven class label distributions.
To ensure a balanced distribution of class labels across datasets, we augment our datasets with two publicly available benchmark datasets, \textbf{GSBench} and \textbf{FCBench} for the GSC and FCP tasks, respectively~\cite{gupta2023learning}.
Using publicly available datasets not only helps maintain a balanced class distribution but also provides the diversity of labelled examples needed to build robust models for the aforementioned tasks~\cite{poolsawad2014balancing}.

\subsection{A.4 Models}
\label{sec:baselines}
The following models are compared in our experiments:


\begin{itemize}[leftmargin=*,align=left]
    \item Rule-Based: 
    simulates the current strategy employed by Support Engineers (SEs), i.e., use predefined rules to detect specific keywords or phrases for log analysis. 
    For GSC and FCP task, we utilize the method proposed in ~\cite{GSC} and ~\cite{FCP}, respectively.
    To the best of our knowledge, no rule-based methods exist for the NER task.

    \item Slate Model~\cite{slatemodel}: A multilingual encoder based transformer model with $153$M parameters ditilled using a XLM-Roberta Base\cite{xlmr}.

    \item Flan-T5-XL~\cite{flan-t5}: An encoder-decoder based transformer model with $3$B parameters from the (Text-to-Text Transfer Transformer) T5 family~\cite{t5transfomer}. We use in-context learning~\cite{icl} to obtain inference predictions by providing few labeled examples in the prompt.
    
    \item BERTOps~\cite{gupta2023learning}: A \textit{logs domain-specific} encoder based transformer model with $110$M parameters continually pre-trained on publicly available log datasets~\cite{loghub}.
\end{itemize}

\textbf{Implementation:} All the experiments were implemented in Pytorch using Hugging Face's Transformers library~\cite{huggingfacelib}.  
The experiments have been conducted on a $A100$ GPU for $200$ epochs, with a weight decay of $0.01$, batch size of $32$ and a learning rate $1e^{-5}$.
Early stopping was applied based on a validation loss, with patience of $10$ evaluations and a threshold of $0.0001$.

\textbf{Evaluation Criteria}
Our log analytics tool is utilized for IT Software Support, where false negatives are a significant concern. 
SEs mustn't overlook any relevant log lines due to the misclassification by the tools underlying models. 
Therefore, we prioritize the \textit{recall} metric, which quantifies a model's ability to accurately identify all relevant instances of a specific class.
Mathematically, recall is defined as: 
\begin{equation}
    \text{Recall}= \frac{True Positive}{True Positive + False Negative}
\end{equation}
To mitigate the impact of class imbalance, \textit{macro-recall} is used as the evaluation metric. 

\begin{table*}[!tbh]
\centering
\small
\begin{tabular}{|c|ccc|c|c|c|ccc|}
\hline
\textbf{\thead{Experiment\\Objective}} & \multicolumn{6}{c|}{\textbf{\thead{Impact of Domain-Specific Few-Shot Examples}}} & \multicolumn{3}{c|}{\textbf{\thead{Generalizability Test using\\Few-Shot Examples}}} \\ \hline
\thead{Training\\Dataset} & \multicolumn{3}{c|}{GSBench} & \thead{GSBench \\+ D1} & \thead{GSBench \\+ D2} & \thead{GSBench \\+ D3} & \multicolumn{3}{c|}{GSBench + D1 + D2 + D3} \\ \hline
\thead{Model/\\Test Dataset} & \multicolumn{1}{c|}{D1} & \multicolumn{1}{c|}{D2} & \multicolumn{1}{c|}{D3} & D1 & D2 & D3 & \multicolumn{1}{c|}{D1} & \multicolumn{1}{c|}{D2} & \multicolumn{1}{c|}{D3} \\ \hline
Rule-Based & \multicolumn{1}{c|}{27.09} & \multicolumn{1}{c|}{24.73} & \multicolumn{1}{c|}{10.4} & 27.09 & 24.73 & 10.4 & \multicolumn{1}{c|}{27.09} & \multicolumn{1}{c|}{24.73} & \multicolumn{1}{c|}{10.4} \\ \hline
Slate Model & \multicolumn{1}{c|}{22.56} & \multicolumn{1}{c|}{17.76} & \multicolumn{1}{c|}{32.89} & 58.24 & 76.08 & 78.2 & \multicolumn{1}{c|}{49.94} & \multicolumn{1}{c|}{61.57} & \multicolumn{1}{c|}{75.63} \\ \hline
Flan-T5-XL & \multicolumn{1}{c|}{26.67} & \multicolumn{1}{c|}{\textbf{39.64}} & \multicolumn{1}{c|}{29.87} & 30.16 & 50.31 & 29.61 & \multicolumn{1}{c|}{38.67} & \multicolumn{1}{c|}{41.47} & \multicolumn{1}{c|}{35.25} \\ \hline
BERTOps & \multicolumn{1}{c|}{\textbf{30.23}} & \multicolumn{1}{c|}{30.08} & \multicolumn{1}{c|}{\textbf{59.72}} & \textbf{72.19} & \textbf{86.31} & \textbf{80.84} & \multicolumn{1}{c|}{\textbf{74.82}} & \multicolumn{1}{c|}{\textbf{82.87}} & \multicolumn{1}{c|}{\textbf{80.84}} \\ \hline
\end{tabular}
\caption{Performance Evaluation on GSC Task}
\label{tab:gs-results-fewshot}

\end{table*}

\begin{table*}[!tbh]
    \centering
    \small
    \begin{tabular}{|c|ccc|c|c|c|c|ccc|}
    \hline
    \textbf{\thead{Experiment\\Objective}} & \multicolumn{6}{c|}{\textbf{\thead{Impact of Domain-Specific Few-Shot Examples}}} & \multicolumn{3}{c|}{\textbf{\thead{Generalizability Test using\\Few-Shot Examples}}} \\ \hline
    \thead{Training\\Dataset} & \multicolumn{3}{c|}{FCBench} & \thead{FCBench \\+ D1} & \thead{FCBench \\+ D2} & \thead{FCBench \\+ D3} & \multicolumn{3}{c|}{FCBench + D1 + D2 + D3} \\ \hline
    \thead{Model/\\Test Dataset} & \multicolumn{1}{c|}{D1} & \multicolumn{1}{c|}{D2} & \multicolumn{1}{c|}{D3} & D1 & D2 & D3 & \multicolumn{1}{c|}{D1} & \multicolumn{1}{c|}{D2} & \multicolumn{1}{c|}{D3} \\ \hline
    Rule-Based & \multicolumn{1}{c|}{35.3} & \multicolumn{1}{c|}{22.45} & \multicolumn{1}{c|}{17.9} & 35.3 & 22.45 & 17.9 & \multicolumn{1}{c|}{35.3} & \multicolumn{1}{c|}{22.45} & \multicolumn{1}{c|}{17.9} \\ \hline
    Slate Model & \multicolumn{1}{c|}{18.57} & \multicolumn{1}{c|}{24.29} & \multicolumn{1}{c|}{18.12} & 72.08 & 70.14 & 77.91 & \multicolumn{1}{c|}{76.31} & \multicolumn{1}{c|}{69.73} & \multicolumn{1}{c|}{86.79}  \\ \hline
    Flan-T5-XL & \multicolumn{1}{c|}{\textbf{60.13}} & \multicolumn{1}{c|}{\textbf{56.31}} & \multicolumn{1}{c|}{\textbf{58.44}} & 65.63 & 60.55 & 58.68 & \multicolumn{1}{c|}{58.63} & \multicolumn{1}{c|}{56.54} & \multicolumn{1}{c|}{50.6}\\ \hline
    BERTOps & \multicolumn{1}{c|}{18.16} & \multicolumn{1}{c|}{20} & \multicolumn{1}{c|}{20} & \textbf{77.68} & \textbf{73.56} & \textbf{89.5} & \multicolumn{1}{c|}{\textbf{78.03}} & \multicolumn{1}{c|}{\textbf{74.53}} & \multicolumn{1}{c|}{\textbf{88.73}}\\ \hline
    \end{tabular}
    \caption{Performance Evaluation on FCP Task}
    \label{tab:fcp-results-fewshot}
    
\end{table*}  

\begin{table}[!tbh]
    \centering
    \resizebox{.48\textwidth}{!}{%
    \begin{tabular}{|c|c|c|c|c|ccc|}
    \hline
    \textbf{\thead{Experiment\\Objective}} & \multicolumn{3}{c|}{\textbf{\thead{Impact of\\Domain-Specific\\Few-Shot Examples}}} & \multicolumn{3}{c|}{\textbf{\thead{Generalizability Test\\using Few-Shot Examples}}} \\ \hline
    \thead{Training\\Dataset} & D1 & D2 & D3 & \multicolumn{3}{c|}{D1 + D2 + D3} \\ \hline
    \thead{Model /\\Test Dataset} & D1 & D2 & D3 & \multicolumn{1}{c|}{D1} & \multicolumn{1}{c|}{D2} & \multicolumn{1}{c|}{D3} \\ \hline
    Rule-Based & - & - & - & - & \multicolumn{1}{c|}{-} & \multicolumn{1}{c|}{-}  \\ \hline
    Slate Model & 94.6 & \textbf{97.67} & 98.25 &  \textbf{98.31} & \multicolumn{1}{c|}{96.18} & \multicolumn{1}{c|}{91.45}\\ \hline
    Flan-T5-XL &  0.016 & 0.009  & 0.0 & \multicolumn{1}{c|}{0.004} & \multicolumn{1}{c|}{0.0016 } & \multicolumn{1}{c|}{0}\\ \hline
    BERTOps & \textbf{98.52} & 97.34 & \textbf{99.1} & \multicolumn{1}{c|}{97.6} & \multicolumn{1}{c|}{\textbf{98.18}} & \multicolumn{1}{c|}{\textbf{93.07}} \\ \hline
    \end{tabular}}
    \caption{Performance Evaluation on NER Task}
    \label{tab:ner-results-fewshot}
    
\end{table}

\subsection{A.5 Evaluation}
In this section, we report the experimental results and present our findings.

\subsubsection{(1) Evaluation in Few-Shot Scenarios}
\label{sec:RQ1}

Tables~\ref{tab:gs-results-fewshot} and ~\ref{tab:fcp-results-fewshot} presents the performance results of models on the GSC and FCP tasks, under two experimental settings: (i) Columns 2-4 display model performance when trained on public data (i.e., GSBench or FCBench) and tested on domain-specific dataset (D1, D2, D3). (ii) Columns 5-7 present the results when models are trained on both public and domain specific datasets (i.e., GSBench + D1 for GSC or FCBench + D1 for FCP) and then tested on respective domain-specific datasets (D1). For instance, the seventh column in Table~\ref{tab:gs-results-fewshot} shows a model trained on GSBench+D3 and tested on D3. 

For both the GSC and FCP tasks, Tables~\ref{tab:gs-results-fewshot} and ~\ref{tab:fcp-results-fewshot} show that adding domain specific examples during training significantly improves the performance of BERTOps and Slate Model.
For example, Slate Model shows a $300\%$ improvement on GSC task when domain-specific data from dataset D2 is provided during training (17.76 in column 3 vs 76.08 in column 6, Table~\ref{tab:gs-results-fewshot}).
Similary, BERTOps achieves a $360\%$ improvement on the FCP task with dataset D1 (from 18.16 in column 2 to 77.68 in column 5, Table~\ref{tab:fcp-results-fewshot}). 

Moreover, BERTOps consistently outperforms other models across both the tasks, likely due to its specialization in the logs domain, which makes it particularly effective in few-shot learning scenarios.
In contrast, Flan-T5-XL's performance remains unchanged despite the inclusion of domain-specific examples. 
This is because it relies on in-context learning rather than fine-tuning. 
Its small context length restricts the number of in-context examples it can utilize, and retraining the model demands significant computational resources and GPUs, making it impractical.
Conversely, both the Slate Model and BERTOps can operate efficiently on CPUs, and consistently outperforms Flan-T5-XL, making them more effective for both GCP and FCP tasks.

Table~\ref{tab:ner-results-fewshot} shows the performance of models on domain-specific datasets for the NER task. 
BERTOps outperforms the others in two out of three datasets and demonstrates comparable performance to the Slate Model on D2. Despite the varying domains of D1-D3, all models perform consistently, suggesting minimal dataset-specific variations in the NER task. 
This consistency indicates that NER is a simpler task compared to GSC and FCP, where models show significant performance differences across datasets.

\begin{tcolorbox}[colback=gray!10, colframe=gray!80, boxrule=0.5pt, left=2pt, right=2pt, top=2pt, bottom=2pt, sharp corners]
\textbf{Key Takeaway:} The results indicate that public annotated datasets often derived from simulated environments, lack the complexity of real-world software support data. Moreover, these datasets may include logs from domains that differ significantly from the intended application domain.
Consequently, training models solely on publicly available datasets may be insufficient for industrial applications; incorporating even a small number of in-domain examples can significantly improve performance.
\end{tcolorbox}

\subsubsection{(2) Generalizability and Robustness}
\label{sec:RQ2}
In this experiment, for each task, a model is finetuned using a training dataset which contains examples from multiple domains (i.e., for GSC: GSBench+D1+D2+D3; for FCP: FCBench+D1+ D2+D3; for NER: D1+D2+D3).

The models are evaluated based on two criteria: 
(1) Generalizability: where, test examples are from the same domain, i.e., (D1, D2 and D3), and,
(2) Robustness: where, test examples are from an unseen domain, i.e., D4. 
The aim of this experiment is to build one model per task which can work across multiple datasets without any changes. 
The last three columns in Tables~\ref{tab:gs-results-fewshot},~\ref{tab:fcp-results-fewshot} and~\ref{tab:ner-results-fewshot} report the generalizability performance of all the models.
For GSC, the results demonstrate that the BERTOps outperforms the Slate Model and Flan-T5-XL by a good margin on all the datasets.
This is because the BERTOps model was continually pretrained on log data\cite{gupta2023learning} whereas the Slate Model or Flan-T5-XL were trained only on the natural language text data.
Interestingly, BERTOps trained GSBench+D1+D2+D3 shows better performance as compared to BERTOps trained solely on GSBench and domain-specific dataset (refer column 5 - 7 in Table~\ref{tab:gs-results-fewshot}). 
For instance, BERTOps trained with GSBench+D1+D2+D3 outperforms BERTOps trained with GSBench+D1 (refer column 8 and 5; $74.82$ vs. $72.19$). Note that the test dataset D1 is the same in both cases.
The improvement in performance can be attributed to the variability introduced during training from examples of different domains\cite{zhu2019aligning}.
However, for D2 test dataset, BERTOps trained on $GSBench+D1+D2+D3$ performs worse than BERTOps trained on GSBench+D2, likely due to overfitting (refer column 6 and 9; 82.87 vs. 86.31).
This is a common issue of poor generalization when models are trained on limited examples~\cite{hiller2022rethinking}. Unlike GSC, for FCP task, Slate Model outperformed its domain-specific variants (refer columns 5 - 10 in Table~\ref{tab:fcp-results-fewshot}).
However, despite this improvement, BERTOps, achieved the best performance on the FCP task as well.
Similar to GSC and FCP, BERTOps outperforms all the other models in the NER task (refer columns 5 - 7 in Table~\ref{tab:ner-results-fewshot}). 
However, in this task, models trained on D1+D2+D3 perform worse than those trained solely on D1, D2 or D3. 
For example, BERTOps's performance decreases from 99.1\% (trained on D3) to 93.07\% (trained on D1+D2+D3). Similarly for Slate Model, 98.25\% (when trained on D3) vs 91.45\% (when trained on D1+D2+D3).
This decline can be attributed to the trade-off in generalizing the model with few shot examples. 
Under such a scenario, it is generally recommended to provide more training examples to help improve the performance.

\begin{table}[!htb]
\small
\centering
\begin{tabular}{|ccc|}
\hline
\multicolumn{3}{|c|}{\textbf{Training}: GSBench + D1 + D2 + D3 ; \textbf{Testing}: D4} \\ \hline
\multicolumn{1}{|c|}{Model/ Scenario} & \multicolumn{1}{c|}{Few Shot} & Full Fine Tuning \\ \hline
\multicolumn{1}{|c|}{Rules-Based} & \multicolumn{1}{c|}{19.64} & 19.64 \\ \hline
\multicolumn{1}{|c|}{Slate Model} & \multicolumn{1}{c|}{29.71} & 54.64 \\ \hline
\multicolumn{1}{|c|}{Flan-T5-XL} & \multicolumn{1}{c|}{28.37} &  27.02 \\ \hline
\multicolumn{1}{|c|}{BERTOps} & \multicolumn{1}{c|}{\textbf{46.59}} & \textbf{66.07} \\ \hline
\end{tabular}
\caption{Performance on Unseen Test Dataset for GSC Task}
\label{tab:gs-results-unseen}
\end{table}

To assess robustness, we focus exclusively on the GSC task. To recap, during training, the models are exposed to examples from GSBench+D1+D2+D3 and tested on a held-out dataset, D4 (not accessible during training).
In this setup, two experimental variations are considered: 
(1) Few-Shot: Models trained from previous experiment (refer columns 8 - 10 in Table~\ref{tab:gs-results-fewshot}). 
(2) Full-Fine Tuning: Model is trained on all the examples from the available datasets (Statistics- GSBench:488, D1:692, D2:755, D3:129). 

Table~\ref{tab:gs-results-unseen} presents the results which indicate that BERTOps outperforms other models by a significant margin, even on unseen datasets. 
In the few-shot setting, BERTOps shows $1.5$x better performance than Flan-T5-XL which performs the worst - likely due to Flan-T5-XL not being fine-tuned for the GSC task and relying solely on in-context learning for inference.
With full fine-tuning, all models (except rule-based) show substantial improvement compared to the few-shot setting. 
For instance, the Slate Model improves approximately by 84\% ($54.64$ vs. $29.71$ pts). 
\begin{tcolorbox}[colback=gray!10, colframe=gray!80, boxrule=0.5pt, left=2pt, right=2pt, top=2pt, bottom=2pt, sharp corners]
\textbf{Key Takeaway:} The results indicate that for all three tasks, BERTOps not only outperforms other models but is also able to generalize well on datasets similar to training datasets' domain. Furthermore, BERTOps performs reasonably well on unseen data.
This insight is crucial as it allows our log analytics tool to scale across multiple products without the need for frequent retraining.
\end{tcolorbox}

\subsection{A.6 Discussion}
From the afforementioned experiments, it is evident that BERTOps outperforms other LLMs such as Flan-T5-XL and Slate Model across the three tasks - GSC, FCP, and NER.
Notably, BERTOps demonstrates superior performance in few-shot learning scenarios and generalizes effectively when trained with examples from multiple domains.
BERTOps also demonstrates strong robustness when tested on unseen datasets, highlighting its effectiveness.
Based on these observations, we chose to integrate BERTOps LLM into our log analytics tool.

\section{B. Production Usage Dashboard}
\begin{figure*}
    \centering
    \includegraphics[width=\textwidth]{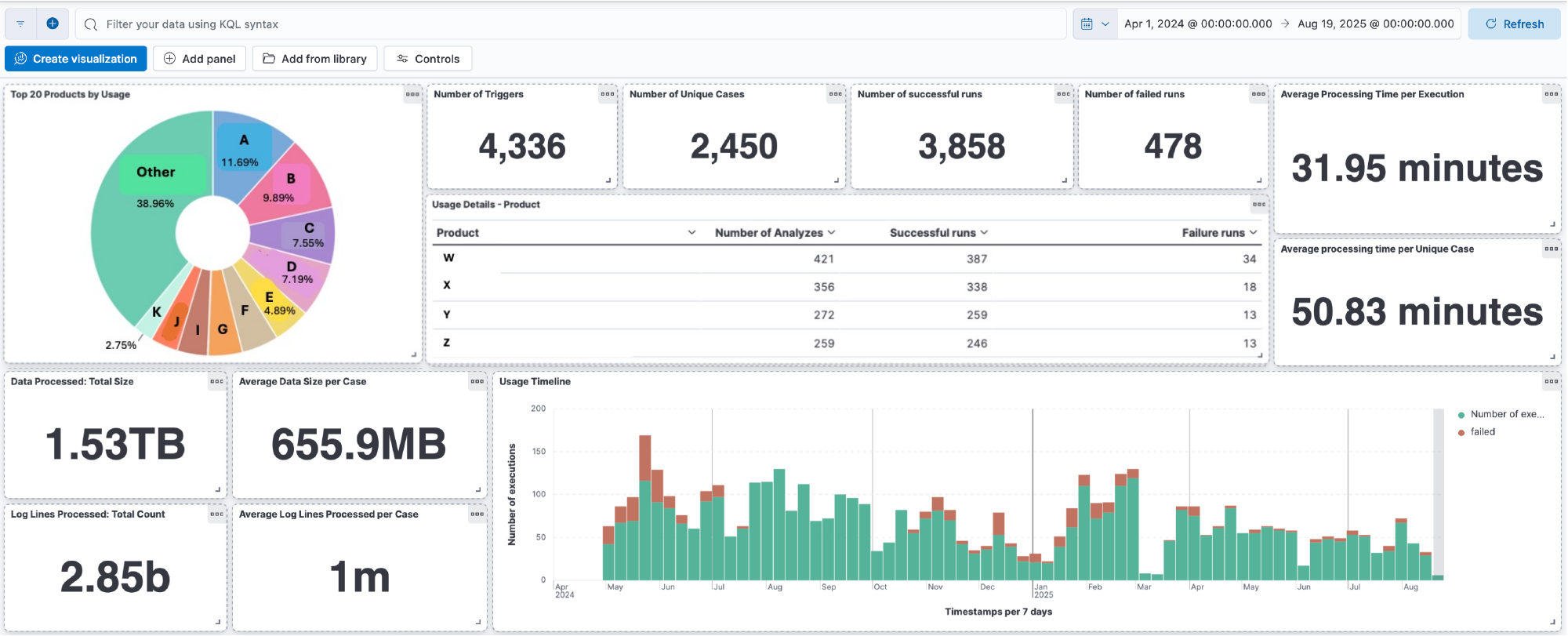}
    \caption{Production Usage Kibana Dashboard of our log analytics tool}
    \label{fig:kibana-dashboard}
\end{figure*}

Figure~\ref{fig:kibana-dashboard} presents a snapshot of the Kibana dashboard used for live monitoring of our log analytics tool, with product names anonymised. Over the past 15 months, the tool has processed more than $2,450$ IT support cases, encompassing approximately $1.53$~TB of log data and over $2.86$~billion log lines. It has been successfully onboarded across $70$ IBM software support products and contributes significant operational savings, reducing costs by more than $\$15,444$ per month and saving $300+$ man hours.

\section{C. User Feedbacks}

Table~\ref{tab:user-feedback} presents real feedback collected from users through the suggestion form at the bottom of the reports in our log analytics tool. Feedback is categorized into three types: (1) Positive Feedback, (2) Negative Feedback, and (3) Feature Requests. The following discussion summarizes the feedback received in each category.

\textbf{Positive Feedback}: Several users highlighted the usefulness of our tool in extracting precise insights through the Golden Signal Classification and Fault Category Prediction components. They appreciated that the tool reduced the cognitive load by generating concise reports, enabling them to quickly identify the potential log lines of interest and to confirm issue timelines. In some cases, the tool enabled users to discover unexpected errors and rare events within the log dump, both of which are very challenging to find manually.

\textbf{Negative Feedback}: Users reported that in some cases, manually reviewing the log dump was simpler than using the tool. Occasionally, the tool's reports showed errors that were not relevant or useful to the problem description. This is acceptable, as our log analytics doesn't currently use the problem description to filter erroneous logs. This filtering capability is planned for future work.
Additionally, some users experienced cases where the tool failed to generate any report even after running for 30 minutes, typically due to unexpected runtime failures, sometimes related to underlying infrastructure, which is beyond our control 

\textbf{Feature Requests}: Users also suggested several enhancement, mostly to improve their user experience. Examples for such requests include, enabling timestamp-based sorting in the reports and provide clear warning messages when unsupported file formats (such as ZIP archives) are used as input. They also emphasized the need to link extracted log lines back to their source files, as well as improving report formatting for multi-line entries like stack traces to improve readability. All of this feedback is being actively monitored and incorporated to improve the tool.

\section{D. Diagnosis Report Screenshot}
\begin{sidewaysfigure*}
    \centering
    \fbox{\includegraphics[width=0.97\textwidth]{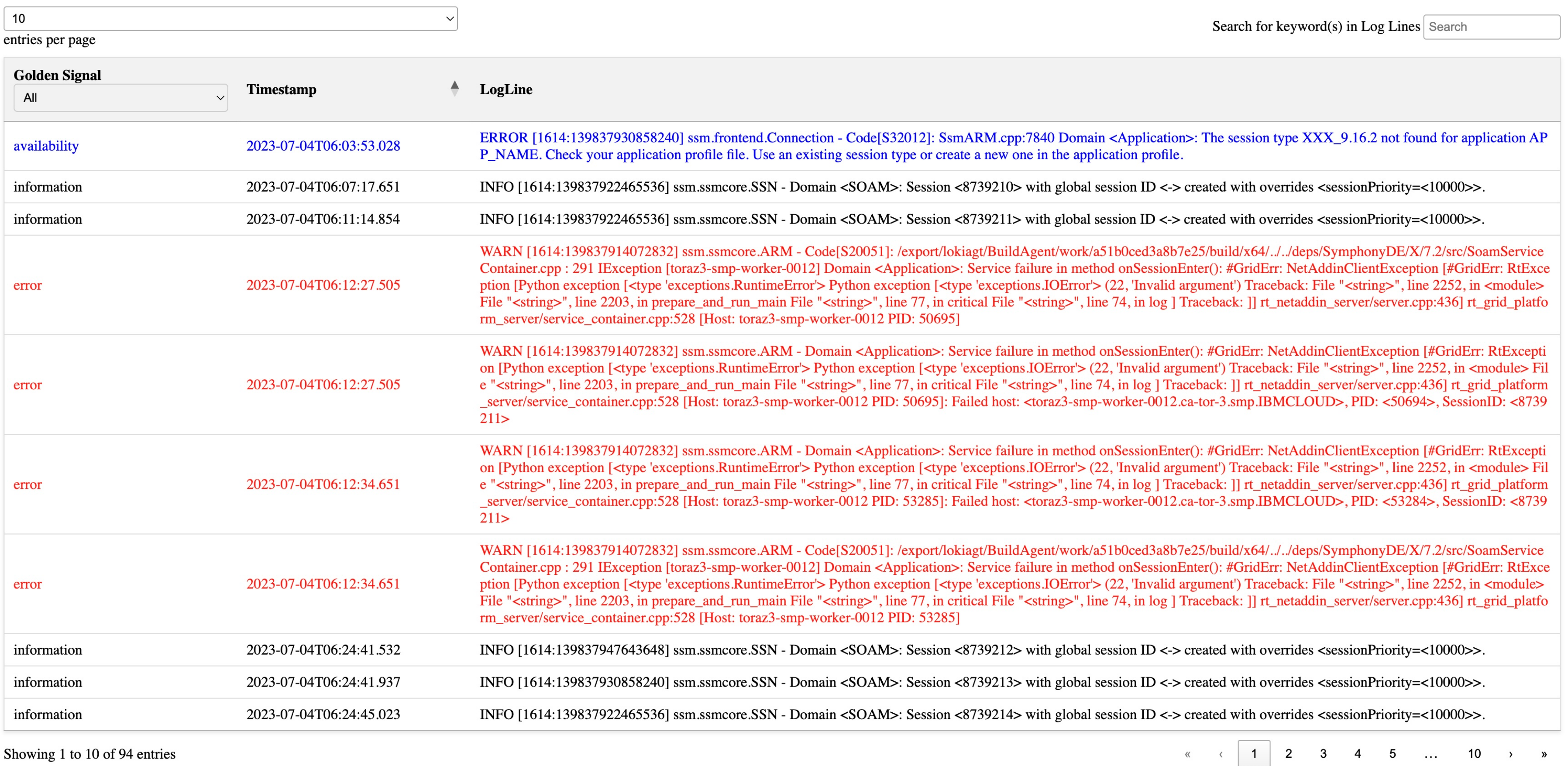}}
    \caption{Diagnosis Report generated by our Log Analytics Tool}
    \label{fig:diagnosis-report}
\end{sidewaysfigure*}

In this section, we present a screenshot of the Diagnosis Report (Figure~\ref{fig:diagnosis-report}) generated by our log analytics tool for a real-world case study from a financial domain application (D4). We include it here due to space constraints in the main content.
The diagnosis reports only shows relevant \textit{log windows} from the log dump, color-coded as per their associated golden signals. In the UI, the log windows are paginated, with each click revealing the next relevant window. A search interface in the top-right corner allows users to filter log lines by keywords.
When combined with insights from the Summary Report, Temporal Trend, and Causal Graph View, the Diagnosis Report enables focused, in-depth analysis of the specific logs identified for investigation.

\begin{table*}[]
\centering
\begin{tabular}{|c|p{14.5cm}|}
\hline
\textbf{Category} & \textbf{Feedback} \\ \hline
\multirow{7}{*}{Positive} & The golden signal classification using the timestamp helped to track the error and the time when the performance issue was reported. Thank you. \\ \cline{2-2}
 & The category of the error feels useful.. whether it is network or application etc.. and I like it when an interpretation is made of an error.. this is harmless etc \\ \cline{2-2}
 & To the point :) \\ \cline{2-2}
 & Was quickly able to identify that there are no logs for the time frame we needed. \\ \cline{2-2}
 & Helped confirm the timelines and some part of the symptoms to look at the logs further. \\ \cline{2-2}
 & Due to issues with the large 2 GB file, I could not save much time on this, but I was curious how many errors of the same time were observed... seems now it was only one! \\ \cline{2-2}
 & In the end, found an unrelated exception that I didn't expect to find, but otherwise the tool was able to break apart a long trace file of SSL data and show there were no failure events in the end. \\ \hline

\multirow{6}{*}{Negative} & Its simpler to manually review the file then use this tool \\ \cline{2-2}
 & I could see where this tool may help in some cases. However, in this case, it basically extracted the error message what was already clearly reported by our own PDCOLLECT file \\ \cline{2-2}
 & It showed an error message, it may not be relevant but it is something else to consider \\ \cline{2-2}
 & It is displaying a lot of useless errors which are not at all relevant to the problem raised in the case. Also took few hours to process 1 mustgather folder. \\ \cline{2-2}
 & No data was returned even after running for 30 min \\ \cline{2-2}
 & I would like to have the line number where the entry was found as usually I'm looking for surrounding messages from what I'm searching for in the filter. \\ \hline

\multirow{5}{*}{Feature Request} & If possible this column: 'Duration (GMT)' is sorted by time? \\ \cline{2-2}
 & The tool has evolved well. It would be great if we add specific timestamp filtering. \\ \cline{2-2}
 & I believe this tool can't run against zip? It should have reported a warning message while running the tool \\ \cline{2-2}
 & Would be helpful to note which files the summary is taking the entries from \\ \cline{2-2}
 & The logline text is wrapped and it is hard to read stack trace because I am unable to identify as where the line starts and ends. It would be of great help if clicking on specific logline takes me to the log file and highlight them. \\ \hline
\end{tabular}
\caption{User Feedback Categorized by Type}
\label{tab:user-feedback}
\end{table*}

\begin{table*}
\centering
\begin{tabular}{|p{10cm}|c|c|c|}
\hline
text & ground-truth & Tool w/o LB & Tool w/ LB \\ \hline
2021-04-23 23:58:41.438 PDT I | \textless PETH\textgreater - Not attempting to disable e1n1.fbond as node is \textbf{UNREACHABLE} (reason: ips is in failed to restart) 
& availability & availability & error \\ \hline
2023-07-17 20:26:44 3539663 [INFO ] ConnectionFetcherImpl.cpp:1221: Test /redfish result: curlcode=0, \textbf{http code=200} 
& information & information & error \\ \hline
\end{tabular}
\caption{Log lines with \textit{important} variables in \textbf{bold} and predictions from Tool w/ LB, Tool w/o LB and Ground Truth.}
\label{tab:inference-output}
\end{table*}

\end{document}